\documentclass{article}
\begin{document}
\font\curs=rsfs10
\def\scri{{\curs I}}

\title{Black Holes and Wormholes in 2+1 Dimensions\thanks{to
appear in Proceedings of 2nd Samos Meeting on Cosmology Geometry and
Relativity, Springer-Verlag}}
\author{Dieter Brill\thanks{email:brill@physics.umd.edu}\\
University of Maryland, College Park, MD 20742, USA
\\and\\ Albert-Einstein-Institut, Potsdam, FRG}
\date{}

\maketitle              
 
\begin{abstract} 
Vacuum Einstein theory in three spacetime dimensions is locally trivial, but admits many
solutions that are globally different, particularly if there is a negative cosmological
constant. The classical theory of such locally ``anti-de Sitter" spaces is treated in
an elementary way, using visualizable models. Among the objects discussed are black holes,
spaces with multiple black holes, their horizon structure, closed universes, and the topologies
that are possible.
\end{abstract} 
 
\section{Introduction} 
 
On general grounds (2+1)-dimensional spacetime was long considered unlikely 
to support black holes, before such solutions were
discovered \cite{BTZ}.
Black holes were commonly conceived as places where the effects of gravity are 
large, surrounded by a region where these effects are asymptotically negligible.
Another possible reason is the idea that black holes are ``frozen
gravitational waves" and therefore exist only in a context where the
gravitational field can have independent degrees of freedom.
In 2+1 dimensional Einstein theory --- that is, Einstein's equations in
a 3-dimensional space-time of signature $(- + +)$ --- the pure,
sourceless gravitational field has no local degrees of freedom,
because in three dimensions the Riemann tensor is given algebraically
by the Einstein tensor, which in turn is algebraically determined by
the Einstein field equations. If there is no matter source and
no cosmological constant, the Riemann tensor vanishes and space-time is
flat; if there is no matter but a cosmological constant $\Lambda$, the
Riemann tensor is that of a space of constant curvature $\Lambda/3$.
Thus gravity does not vary from place to place and it does not have
any wave degrees of freedom. These were some of the reasons
why the possibility of black holes was discounted, and the discovery of
black hole solutions in 2+1 D spacetimes with a negative $\Lambda$ came as
such a surprise.

The existence of (2+1)-dimensional black holes of course does not alter the 
absence of gravitational waves in (2+1)-dimensional Einstein spaces, nor the 
lack of variation of their curvature. The curvature of spacetimes 
satisfying the sourceless Einstein equations with negative $\Lambda$ is 
constant negative, and the local geometry in the asymptotic region does not 
differ from that near the black hole. Indeed, black hole solutions 
can be obtained from 
the standard, simply connected spacetime of constant negative curvature (anti-de 
Sitter space, AdS space for short) by forming its quotient space with a 
suitable group of 
isometries.\footnote{It appears that all locally AdS spacetimes can be obtained in 
this way \cite{mess}. This is not so for
positive curvature \cite{MJW}.} One of
the criteria on the isometries is that the quotient space should not have any
objectionable singularities. For example, if the group contains 
isometries of the rotation type, with a timelike set of fixed points,
then the quotient space will have singularities of the conical kind.
Such singularities can represent ``point" particles, and the corresponding
spacetime can be interpreted as an interesting and physically meaningful
description of the dynamics of such particles \cite{Hoo}. However, we confine 
attention to solutions of the sourceless Einstein equations with negative 
cosmological constant --- whether black holes or not --- that  are at least 
initially nonsingular. Therefore we exclude such particle-like solutions. 
(Likewise, we will not consider the interesting developments in 
lower-dimensional dilaton gravity, nor other matter fields \cite{LD}.)

On the other hand, the group used to construct our quotient space may have 
isometries that are locally Lorentz boosts, with spacelike sets of 
fixed points. The 
corresponding singularities are of the non-Hausdorff ``Misner" type \cite{Mis}. If 
such a singularity does not occur on an initial spacelike surface, and is hidden 
behind an event horizon, then the spacetime can be acceptable as a 
representation of a black hole. Finally, the isometry may not have any fixed 
points but still lead to regions in the quotient space that are to be considered 
singular for physical reasons, and such regions may again be surrounded by an 
event horizon, yielding other types of black holes.

Thus the proper criterion characterizing a black hole in this context is not 
a region of large curvature or an infinite red shift (in typical representations 
of AdS space itself, where there is no black hole, there is an infinite red 
shift between the interior and the region near infinity), but existence of an 
event horizon. This in turn requires the existence of a suitable \scri, whose 
neighborhood is a region in which ``distant observers" can survive for an 
arbitrarily long time without hitting a singularity. That is, there have to be 
causal curves (the worldlines of these observers) that can be continued to 
infinite proper time. For example, Misner space itself --- the quotient of 
Minkowski space by a Lorentz boost --- does not satisfy this criterion in 
any dimension, because all timelike curves intersect the non-Hausdorff 
singularity in a finite proper time. Thus the case $\Lambda = 0$ does not yield 
any black holes. The same is true, for similar reason, in the case $\Lambda > 
0$. However, for $\Lambda < 0$ there are worldlines along which asymptotic
observers can survive forever even when spacelike singularities are present. 
Our black holes will then not be asymptotically flat \cite{Heu}, but
asymptotically AdS.
We will see (in section 3) that the usual definition of black holes can be
applied to these spacetimes, and even before we have come to this 
we will speak of them as black holes. 

We can understand the difference between the cases $\Lambda \geq 0$ and
$\Lambda < 0$ as as a consequence of the positive 
``relative acceleration" of spacelike geodesics in spaces of negative 
curvature. Spacelike geodesics reaching the asymptotically AdS region 
will increase their separation without limit. The 
fixed points of the identification that generates a black hole --- that 
is, the ``singularities" --- lie along a spacelike geodesic. Consider a 
set of observers located initially further and further towards the
asymptotic region and along another spacelike geodesic, which does not 
intersect the geodesic of fixed points. The timelike distance of an 
observer from the singularity 
will then eventually increase without limit, so a sufficiently far-out 
observer can survive for an arbitrarily long time.

We note in passing that {\em timelike} geodesics in spacetimes of constant 
negative curvature have the opposite property: they accelerate toward 
each other. Thus $\Lambda < 0$ corresponds to a universal ``attractive"
gravity, and a black hole in such a spacetime exerts this same attraction 
on test particles, as a black hole should.

The quotient of the AdS universe with the group generated by a single 
finite isometry that is without fixed points, at least on some initial 
spacelike surface, yields a single black hole, called a BTZ spacetime 
(for its discoverers, Ba\~nados, Teitelboim and Zanelli \cite{BTZ}). 
As we will see, one can make further identifications in a BTZ spacetime, 
obtaining more complicated black holes, and this process can be repeated 
an arbitrary number of times. Although the isometries used for 
the identification cannot be entirely 
arbitrary, the variety of possibilities and of the resulting spacetimes 
is quite large. These spacetimes cannot be described by their metric in 
one or in a few simple coordinate systems, because many coordinate patches 
would be needed to cover their possibly complicated topology. In 
principle such a spacetime is of course defined, and all its physical 
properties are computable, once we know the structure of the AdS 
isometries that generate it. But such a presentation does not give an 
accessible and easily visualizable picture of the spacetime. Therefore 
we prefer to describe the spacetimes combinatorially, by ``gluing together" 
pieces of AdS space. This view allows one to gain many important
geometrical insights directly, without much algebra or analysis (even if 
a few of these geometrical constructions may resemble a tour de force).

In section 2 we consider the simplest, time-symmetric case. Because the 
extrinsic curvature of the surface of time-symmetry vanishes, this 
surface is itself a smooth two-dimensional Riemannian space of constant 
negative curvature. This class of spaces
has been studied in considerable detail \cite{BP}. In particular, almost 
all two-dimensional spacelike topologies occur already within this class. 
Section 3 considers the time development of these spaces; we find that all the 
non-compact initial states  develop into black holes.  The horizon can 
be found explicitly, although its behavior can be quite complicated.
Section 4 concerns spacetimes that are not time-symmetric but have angular 
momentum.

An important reason for studying the classical behavior of these spacetimes is 
their relative simplicity while still preserving many of the features of more 
realistic black hole spacetimes. They are therefore interesting models for 
testing the formalism of quantum gravity. We do not go into these developments 
but refer the reader to the recent book by S. Carlip \cite{Carl}.

\section{Time-symmetric geometries}

Three-dimensional AdS space has many totally geodesic (``time-symmetric") 
spacelike 
surfaces. Because the extrinsic curvature of such surfaces vanishes, they 
have constant negative curvature $\Lambda$. Each such surface remains
invariant under a
``little group" of AdS isometries, which are therefore isometries of the 
spacelike surface, and conversely each isometry of the spacelike surface 
can be extended to be an isometry of the whole AdS 
spacetime.\footnote{Since AdS spacetime is an analytic continuation (both 
in signature and curvature) of the familiar spherical geometry, such 
properties can be considered extensions of the corresponding statements 
about spheres, mutatis mutandis for the difference in group structure,
SO(4) vs SO(2,2). Analogous statements are true about surfaces of 
constant extrinsic curvature.} Therefore any identification obtained by 
isometries on the spacelike surface can likewise be extended to the whole 
spacetime. (AdS space identified by this extension coincides with the usual 
time development of the initial data via Einstein's equations where the 
latter is defined, but it even goes beyond any Cauchy horizon). 
Thus to identify the possible time-symmetric geometries it suffices to discuss the 
possible initial spacelike geometries --- although this leaves the time 
development still to be made explicit.

\subsection{Coordinates} 

Although most physically and mathematically interesting facts about 
constant negative curvature spaces can be phrased without 
reference to coordinates, and even usefully so, it is convenient for the 
elucidation and proof of these facts to have coordinates available. 
Because of the large number of symmetries of AdS spacetime, its geometry 
takes a simple form in a large number of coordinate systems, which do not 
usually cover all of the spacetime, but which exhibit explicitly one or 
several of these symmetries. The simplest coordinates are the redundant 
set of four $X^\mu$, $\mu = 1,\dots 4$ in terms of which AdS space is 
usually defined, namely as an embedding in four-dimensional flat space 
with signature $(-,-,+,+)$ and metric
\begin{eqnarray}
   ds^2=-dU^2-dV^2+dX^2+dY^2
   \label{emme}
\end{eqnarray}
by the surface
\begin{eqnarray}
   -U^2-V^2+X^2+Y^2=-\ell^2 .
   \label{ads}
\end{eqnarray}
This spacetime is periodic in the timelike direction with
the topology $S^1\times R^2$; for example, for $X^2+Y^2<\ell^2$
the curves ($X,\,Y$) = const, 
$ U^2+V^2 = \ell^2 - X^2-Y^2$ are closed timelike circles. 
In the following we assume that this periodicity has been removed by passing to the 
universal covering space with topology $R^3$, which we will call AdS 
space. If it is necessary to distinguish it from the space of Eq 
(\ref{ads}) we will call the latter ``periodic AdS space." Either 
spacetime is a solution of the vacuum Einstein 
equations with a negative cosmological constant $\Lambda=-1/\ell^2$.

Eq (\ref{ads}) shows that AdS space is a surface of constant distance 
from the origin in the metric (\ref{emme}). It therefore inherits from the 
embedding space all the isometries that leave the origin fixed, which 
form the SO(2,2) group. AdS space can be described by coordinates 
analogous to the usual spherical polar coordinates as in Eq (\ref{ts}), but
of greater interest are coordinates related to 
isometries that leave a plane fixed, and whose orbits lie in the orthogonal 
plane.  These have the nature of rotations if the plane is spacelike (or double-timelike, 
such as the ($U,\,V$) plane), and of Lorentz transformations if the plane is timelike.
Isometries corresponding to orthogonal planes commute, and we can find coordinates that 
exhibit such pairs of isometries explicitly. If the isometries are rotations, the 
coordinates cover all of AdS space; if they are Lorentz transformations the 
corresponding 
coordinates are analogous to Rindler coordinates of flat space, and need to be 
analytically extended in the usual fashion to cover all of the spacetime.

For example, if we choose rotations by an angle $\theta$ in the ($X,\,Y$) plane and by an 
angle $t/\ell$ in the ($U,\,V$) plane, and specify the respective orbits 
on the AdS surface by
$$U^2+V^2=\ell^2\cosh^2\chi \qquad {\rm and} \qquad X^2+Y^2 = \ell^2\sinh^2\chi$$
(so that,  for example, $U=-\ell\cosh\chi\cos{t\over\ell},\;
V=\ell\cosh\chi\sin{t\over\ell}$) we obtain the metric
\begin{eqnarray}
   ds^2=-\cosh^2\chi dt^2+\ell^2\left( d\chi^2+\sinh^2\chi d\theta^2\right) .
   \label{statmet}
\end{eqnarray}
In order to describe the universal covering space we have to allow $t$ to 
range
$-\infty < t < \infty$, whereas $\theta$ has its usual range, $0\leq\theta<2\pi$,
and similarly $0\leq\chi<\infty$. Except for the usual polar coordinate 
singularity at $\chi = 0$, these coordinates cover all of AdS space by a 
sequence of identical (``static") two-dimensional spacelike surfaces 
$t =$ const having a standard metric of spaces of constant negative curvature 
$-1/\ell^2$. Because $U=0=V$ does not occur on (\ref{ads}),
shifts in the $t$ coordinate are true translations, without fixed points. 
These coordinates define timelike sections ($\theta=$ const) and spacelike 
sections ($t=$ const) of AdS space. Each of these 
can be represented in a conformal diagram, shown in Fig.~1. 

\begin{figure}
\unitlength 0.8mm
\linethickness{0.4pt}
\begin{picture}(140.20,86.00)(10,5)
\bezier{184}(99.80,50.00)(99.80,61.28)(109.90,67.68)
\bezier{180}(109.90,67.68)(120.00,72.73)(130.10,67.68)
\bezier{184}(130.10,67.68)(140.20,61.28)(140.20,50.00)
\bezier{184}(99.80,50.00)(99.80,38.72)(109.90,32.32)
\bezier{180}(109.90,32.32)(120.00,27.27)(130.10,32.32)
\bezier{184}(130.10,32.32)(140.20,38.72)(140.20,50.00)
\put(20.00,15.00){\line(0,1){70.00}}
\put(40.00,15.00){\line(0,1){70.00}}
\put(60.00,15.00){\line(0,1){70.00}}
\put(80.00,50.00){\makebox(0,0)[cc]{$t=0$}}
\put(85.00,50.00){\vector(1,0){11.00}}
\put(75.00,50.00){\vector(-1,0){11.00}}
\put(20.00,50.00){\line(1,0){40.00}}
\put(20.00,60.00){\line(1,0){40.00}}
\put(20.00,70.00){\line(1,0){40.00}}
\put(20.00,80.00){\line(1,0){40.00}}
\put(20.00,40.00){\line(1,0){40.00}}
\put(20.00,30.00){\line(1,0){40.00}}
\put(20.00,20.00){\line(1,0){40.00}}
\put(48.00,82.00){\line(0,-1){64.00}}
\put(54.00,82.00){\line(0,-1){64.00}}
\put(57.00,82.00){\line(0,-1){64.00}}
\put(32.00,82.00){\line(0,-1){64.00}}
\put(26.00,82.00){\line(0,-1){64.00}}
\put(23.00,82.00){\line(0,-1){64.00}}
\bezier{184}(103.38,50.00)(103.38,59.28)(111.69,64.54)
\bezier{180}(111.69,64.54)(120.00,68.70)(128.31,64.54)
\bezier{184}(128.31,64.54)(136.62,59.28)(136.62,50.00)
\bezier{184}(103.38,50.00)(103.38,40.72)(111.69,35.46)
\bezier{180}(111.69,35.46)(120.00,31.30)(128.31,35.46)
\bezier{184}(128.31,35.46)(136.62,40.72)(136.62,50.00)
\bezier{184}(106.33,50.00)(106.33,57.63)(113.16,61.97)
\bezier{180}(113.16,61.97)(120.00,65.38)(126.84,61.97)
\bezier{184}(126.84,61.97)(133.67,57.63)(133.67,50.00)
\bezier{184}(106.33,50.00)(106.33,42.37)(113.16,38.03)
\bezier{180}(113.16,38.03)(120.00,34.62)(126.84,38.03)
\bezier{184}(126.84,38.03)(133.67,42.37)(133.67,50.00)
\bezier{184}(112.01,50.00)(112.01,54.46)(116.00,57.00)
\bezier{180}(116.00,57.00)(120.00,58.99)(124.00,57.00)
\bezier{184}(124.00,57.00)(127.99,54.46)(127.99,50.00)
\bezier{184}(112.01,50.00)(112.01,45.54)(116.00,43.00)
\bezier{180}(116.00,43.00)(120.00,41.01)(124.00,43.00)
\bezier{184}(124.00,43.00)(127.99,45.54)(127.99,50.00)
\put(100.00,50.00){\line(1,0){40.00}}
\put(120.00,30.00){\line(0,1){40.00}}
\put(138.00,61.67){\makebox(0,0)[lb]{$\theta$}}
\put(119.33,49.67){\vector(3,1){0.2}}
\bezier{100}(99.33,37.00)(107.33,46.00)(119.33,49.67) 
\put(99.00,36.33){\makebox(0,0)[ct]{$\chi = 0$}}
\put(105.67,35.67){\line(1,1){29.00}}
\put(134.33,35.67){\line(-1,1){29.00}}
\put(140.00,42.00){\makebox(0,0)[lt]{$\chi=\infty$}}  
\put(40.00,86.00){\makebox(0,0)[cb]{$\chi=0$}}
\put(60.00,86.00){\makebox(0,0)[cb]{$\chi=\infty$}}   
\put(61.00,70.00){\makebox(0,0)[lc]{$t=$ const}}
\put(40.00,10.00){\makebox(0,0)[cc]{({\bf a})}}
\put(120.00,10.00){\makebox(0,0)[cc]{({\bf b})}}
\thicklines
\put(135.67,62.67){\vector(-1,1){0.20}}
\end{picture}
\caption{Conformal diagrams of the static (or sausage) coordinates 
of Eq (\ref{statmet}) 
in sections of AdS space. 
({\bf a}) The $\chi,\,t$ section, both sides of the origin. The right half is, for 
example, $\theta = 0$, and the left half, $\theta=\pi$. ({\bf b}) The section $t=$ const is 
the 2D space of constant negative curvature, conformally represented as a Poincar\'e disk 
(see section 2.2). The conformal factors are different in the two sections, so they do not 
represent sections of one three-dimensional conformal diagram. (For the 
latter see Fig.~4b)} \label{fig1} 
\end{figure}
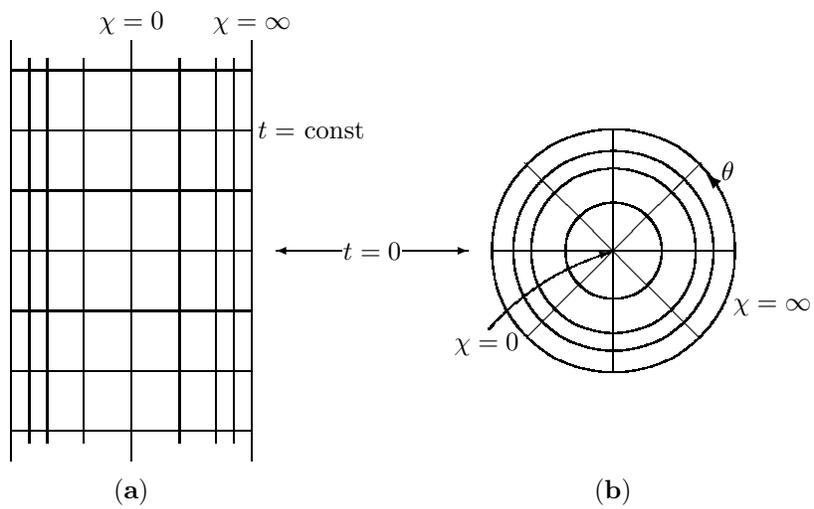  

We can define a ``radial" coordinate (which really measures the circumference of circles) by
$$ r = \ell\sinh\chi\,.$$
The metric (\ref{statmet}) then takes the form
\begin{equation}
   ds^2=-\left({r^2\over\ell^2}+1\right) dt^2+
   \left({r^2\over\ell^2}+1\right)^{-1} dr^2+r^2 d\theta^2\,.
	\label{m0}
\end{equation}
By choosing a different radial coordinate, namely $$\rho = \ell\tanh{1\over 2}\chi$$ 
to replace the $\chi$ of Eq (\ref{statmet}), we can make the conformally flat nature of the 
spacelike section explicit and keep the metric static: 
\begin{equation}
	ds^2=-\left({1+(\rho/\ell)^2\over 1-(\rho/\ell)^2}\right)^2dt^2 +
		{4\over\left(1-(\rho/\ell)^2\right)^2}\left(d\rho^2 + \rho^2d\theta^2\right).
	\label{sausge}
\end{equation}

A picture like Fig.~1, with parts ({\bf a}) and ({\bf b}) put together into a 
3-dimensional cylinder, can be considered a plot of AdS space in the cylindrical 
coordinates of Eq (\ref{sausge}).
Because of the cylindrical shape of this diagram these coordinates are sometimes called {\it 
sausage coordinates} \cite{Ingemar}. Like the static coordinates of (\ref{statmet}), these 
cover all of AdS space.

If we follow an analogous construction but use the timelike ($X,\, U$) and ($Y,\, V$) planes 
with orbits (in terms of a new coordinate $\chi$)
$$-V^2+X^2=-\ell^2\cosh^2\chi \qquad {\rm and} \qquad -U^2+Y^2 = \ell^2\sinh^2\chi ,$$
and new hyperbolic coordinates $\phi$ and $t/\ell$, we obtain the metric 
\begin{equation}
ds^2 = -\sinh^2\chi dt^2 +\ell^2\left(d\chi^2+\cosh^2\chi d\phi^2\right).
\end{equation}
By defining
$$r = \ell\cosh\chi$$
we can change this to the Schwarzschild-coordinate form
\begin{eqnarray} 
   ds^2=-\left({r^2\over\ell^2}-1\right) dt^2+
   \left({r^2\over\ell^2}-1\right)^{-1} dr^2+
   r^2 d\phi^2,
   \label{schwmet}
\end{eqnarray}  
which is usually derived from the ``rotationally" symmetric ansatz --- however, in this 
description 
of AdS space, $\phi$ has to be given the full range, $-\infty < \phi < \infty$ of a 
hyperbolic angle. The range of $r$ for which the metric (\ref{schwmet}) is regular, 
$\ell<r<\infty$, 
describes only a part of AdS space, as can be seen from the explicit 
expression for the embedding in terms of these coordinates, 
\begin{eqnarray}
   \left\{\begin{array}{l}
   U=\left(r^2-\ell^2\right)^{1/2}\sinh{t\over\ell}\\
   V=r\cosh\phi\\
   X=r\sinh\phi\\
   Y=\left(r^2-\ell^2\right)^{1/2}\cosh{t\over\ell}\\   
   \end{array}
   \right.
   \label{schwc}
\end{eqnarray}
This regular region can be patched together in the usual way with the 
region $0<r<\ell$ (Fig.~2), to describe a larger part of AdS space. But 
if it is desired (for whatever bizarre reason) to describe all of AdS space 
by analytic extensions of the coordinates (\ref{schwc}), one needs also 
analytic extensions beyond the null surfaces $\phi=\pm\infty$ (or $r=0$), 
which are quite analogous to the usual Schwarzschild-type ``horizon" null 
surfaces $t=\pm\infty$ (or $r=\ell$). One then finds two disjoint regions 
of a third type (not shown in the Figure because they extend perpendicular 
to the plane of Fig.~2a) in which $r^2$ is negative and $\phi$ is the timelike 
coordinate.\footnote{Like all statements derived from embedding equations 
such as (\ref{schwc}) this really applies to periodic AdS space, and 
should be repeated an infinite number of times for the covering AdS space 
itself. For example, there are an infinite number of regions of the three 
types in AdS space.}

\begin{figure}
\unitlength 0.80mm
\linethickness{0.4pt}
\begin{picture}(131.00,90.00)
\bezier{184}(89.80,50.32)(89.80,61.60)(99.90,68.00)
\bezier{180}(99.90,68.00)(110.00,73.05)(120.10,68.00)
\bezier{184}(120.10,68.00)(130.20,61.60)(130.20,50.32)
\bezier{184}(89.80,50.32)(89.80,39.05)(99.90,32.64)
\bezier{180}(99.90,32.64)(110.00,27.59)(120.10,32.64)
\bezier{184}(120.10,32.64)(130.20,39.05)(130.20,50.32)
\put(10.00,10.00){\line(0,1){80.00}}
\put(50.00,10.00){\line(0,1){80.00}}
\put(10.00,50.00){\line(1,0){40.00}}
\put(10.00,30.00){\line(1,1){40.00}}
\put(50.00,30.00){\line(-1,1){40.00}}
\bezier{88}(10.00,42.00)(18.00,42.00)(30.00,50.00)
\bezier{88}(50.00,58.00)(42.00,58.00)(30.00,50.00)
\bezier{88}(10.00,58.00)(18.00,58.00)(30.00,50.00)
\bezier{88}(50.00,42.00)(42.00,42.00)(30.00,50.00)
\put(10.00,70.00){\line(1,0){40.00}}
\bezier{204}(50.00,70.00)(34.00,50.00)(50.00,30.00)
\bezier{204}(10.00,70.00)(26.00,50.00)(10.00,30.00)
\bezier{204}(50.00,70.00)(30.00,54.00)(10.00,70.00)
\put(30.00,5.00){\makebox(0,0)[cc]{({\bf a})}}
\put(90.00,50.00){\line(1,0){40.00}}
\bezier{152}(95.00,64.00)(110.00,52.00)(125.00,64.00)
\bezier{40}(102.00,69.00)(104.00,64.00)(110.00,64.00)
\bezier{40}(118.00,69.00)(116.00,64.00)(110.00,64.00)
\bezier{152}(95.00,37.00)(110.00,49.00)(125.00,37.00)
\bezier{40}(102.00,32.00)(104.00,37.00)(110.00,37.00)
\bezier{40}(118.00,32.00)(116.00,37.00)(110.00,37.00)
\put(10.00,30.00){\line(1,0){40.00}}
\bezier{204}(50.00,30.00)(30.00,46.00)(10.00,30.00)
\bezier{216}(110.00,30.00)(92.00,50.00)(110.00,70.50)
\bezier{120}(110.00,30.00)(95.00,36.00)(95.00,50.00)
\bezier{120}(110.00,70.50)(95.00,64.00)(95.00,50.00)
\bezier{216}(110.00,30.00)(128.00,50.00)(110.00,70.50)
\bezier{120}(110.00,30.00)(125.00,36.00)(125.00,50.00)
\bezier{120}(110.00,70.50)(125.00,64.00)(125.00,50.00)
\put(70.00,50.00){\makebox(0,0)[cc]{$t=0$}}
\put(75.00,50.00){\vector(1,0){12.00}}
\put(65.00,50.00){\vector(-1,0){12.00}}
\put(51.00,58.00){\makebox(0,0)[lc]{$t=$ const}}   
\put(30.00,70.33){\makebox(0,0)[cb]{$r=0$}}
\put(56.00,75.00){\makebox(0,0)[lc]{$r=$ const}}
\put(39.00,64.00){\vector(-1,-2){0.2}}
\bezier{92}(55.00,75.00)(43.00,74.33)(39.00,64.00)
\put(44.00,60.00){\vector(-3,-1){0.2}}
\bezier{80}(56.00,73.00)(54.00,63.33)(44.00,60.00)
\put(110.00,55.00){\makebox(0,0)[cc]{$r$=$\ell$}} 
\put(131.00,50.00){\makebox(0,0)[lc]{$\phi=0$}} 
\put(129.00,62.00){\makebox(0,0)[lc]{$\phi$}}   
\put(110.00,5.00){\makebox(0,0)[cc]{({\bf b})}}
\put(110.00,30.00){\line(0,1){23.00}}
\put(110.00,56.00){\line(0,1){14.50}}
\put(110.00,71.00){\makebox(0,0)[cb]{$\phi=\infty$}}
\put(110.00,29.00){\makebox(0,0)[cc]{$\phi=-\infty$}}
\thicklines
\put(127.00,62.00){\vector(-1,2){0.20}}
\bezier{10}(10.00,50.00)(20.00,60.00)(30.00,70.00)
\bezier{10}(30.00,70.00)(40.00,60.00)(50.00,50.00)
\bezier{10}(50.00,50.00)(40.00,40.00)(30.00,30.00)
\bezier{10}(30.00,30.00)(20.00,40.00)(10.00,50.00)
\bezier{20}(10.00,50.00)(30.00,65.00)(50.00,50.00)
\bezier{20}(50.00,50.00)(30.00,35.00)(10.00,50.00)
\end{picture}
\caption{Conformal diagrams of the ``Schwarzschild" coordinates of 
Eq (\ref{schwc}) in sections of AdS space. 
({\bf a})~An $r,\,t$ section, continued across the $r=\ell$ coordinate singularity. 
The outer vertical lines correspond to $r=\infty$. The dotted curves show a
few of the surfaces $\tau=$ const for the coordinates of Eq (\ref{cmc}),
with limits at $\tau = \pm\pi\ell/2$. 
({\bf b})~An $r,\,\phi$ section ($r>\ell$) is 
a two dimensional space of constant negative curvature, conformally represented as a 
Poincar\'e disk (see below). The approximately vertical curves are lines of constant $r$; 
they are equidistant in the hyperbolic metric. The approximately horizontal curves are 
lines of constant $\phi$; they are geodesics in the hyperbolic metric. The outer circle 
corresponds to $r=\infty$} 
\end{figure}
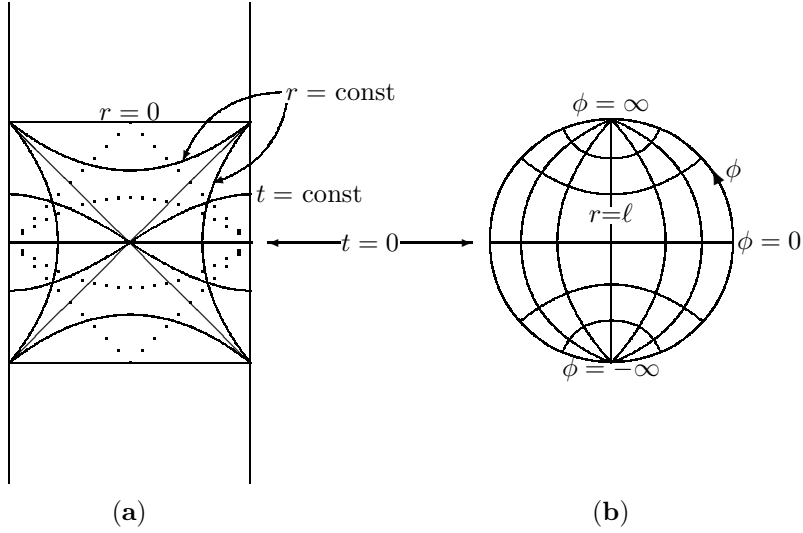  

Another interesting coordinate system is closely related to ordinary polar
coordinates on the three-sphere:
\begin{eqnarray}
   \left\{\begin{array}{l}
   U=\ell\sin{\tau\over\ell}\\
   V=\left(r^2-\ell^2\right)^{1/2}\cos{\tau\over\ell}\\
   X=r\cos{\tau\over\ell} \cos\phi \\
   Y=r\cos{\tau\over\ell} \sin\phi\\   
   \end{array}
   \right.
   \label{ts}
\end{eqnarray}
with the metric
\begin{equation}
	ds^2=-d\tau^2 + \cos^2\left({\tau\over\ell}\right)
	\left[\left({r^2\over\ell^2}-1\right)^{-1}dr^2+ r^2 d\phi^2\right].
	\label{cmc}
\end{equation}
This is a time development of the same initial data as in (\ref{schwmet}) (at $t=0$
resp.\ $\tau=0$) but with unit lapse function $N=1$. The surfaces $\tau=$ const have
constant extrinsic curvature, and they just cover the domain of dependence of
those initial values.  

Finally one can introduce coordinates that correspond to the flat sections of
de Sitter space:
\begin{eqnarray}
   \left\{\begin{array}{l}
   U+Y=r\\
   U-Y=r\left(\phi^2-t^2\right)+{1\over r}\\
   X=r\phi\\
   V=rt\\   
   \end{array}
   \right.
   \label{horo}
\end{eqnarray}
The metric then takes the form
\begin{equation}
	ds^2=-r^2dt^2+{dr^2\over r^2}+r^2d\phi^2.
	\label{horom}
\end{equation}
Here the $r=$ const sections are manifestly flat.\footnote{These subspaces 
are the analog in the case of Lorentzian metrics of {\it horospheres} of 
hyperbolic spaces (see, for example, \cite{BP}).} Fig.~3 shows the conformal 
picture of these coordinates.

\begin{figure}
\unitlength 0.80mm
\linethickness{0.4pt}
\begin{picture}(131.00,90.00)(5,0)
\bezier{184}(89.80,50.32)(89.80,61.60)(99.90,68.00)
\bezier{180}(99.90,68.00)(110.00,73.05)(120.10,68.00)
\bezier{184}(120.10,68.00)(130.20,61.60)(130.20,50.32)
\bezier{184}(89.80,50.32)(89.80,39.05)(99.90,32.64)
\bezier{180}(99.90,32.64)(110.00,27.59)(120.10,32.64)
\bezier{184}(120.10,32.64)(130.20,39.05)(130.20,50.32)
\put(10.00,10.00){\line(0,1){80.00}}
\put(50.00,10.00){\line(0,1){80.00}}
\put(30.00,5.00){\makebox(0,0)[cc]{({\bf a})}}
\put(70.00,50.00){\makebox(0,0)[cc]{$t=0$}}
\put(75.00,50.00){\vector(1,0){12.00}}
\put(65.00,50.00){\vector(-1,0){12.00}}
\put(131.00,50.00){\makebox(0,0)[lc]{$\phi=0$}}
\put(127.00,61.50){\vector(-2,3){1.00}}
\put(127.67,63.00){\makebox(0,0)[lc]{$\phi$}}
\put(110.00,5.00){\makebox(0,0)[cc]{({\bf b})}}
\put(10.00,50.00){\line(1,1){40.00}}
\put(10.00,50.00){\line(1,-1){40.00}}
\put(10.00,50.00){\line(1,0){40.00}}
\put(90.00,50.00){\line(1,0){40.00}}
\bezier{144}(90.00,50.00)(108.00,52.00)(110.00,70.00)
\bezier{144}(90.00,50.00)(108.00,48.00)(110.00,30.00)
\bezier{160}(90.00,50.00)(107.00,49.00)(124.00,64.00)
\bezier{160}(90.00,50.00)(107.00,51.00)(124.00,36.00) 
\bezier{120}(89.66,50.22)(89.66,57.85)(96.49,62.18)
\bezier{120}(96.49,62.18)(103.33,65.60)(110.17,62.18)
\bezier{120}(110.17,62.18)(117.00,57.85)(117.00,50.22)
\bezier{120}(89.66,50.22)(89.66,42.59)(96.49,38.25)
\bezier{120}(96.49,38.25)(103.33,34.83)(110.17,38.25)
\bezier{120}(110.17,38.25)(117.00,42.59)(117.00,50.22)
\put(96.00,50.00){\circle{12.00}}
\bezier{380}(50.00,90.00)(24.00,50.00)(50.00,10.00)
\bezier{204}(22.00,50.00)(22.00,58.00)(50.00,90.00)
\bezier{204}(22.00,50.00)(22.00,42.00)(50.00,10.00)
\bezier{180}(10.00,50.00)(36.00,66.00)(50.00,66.00)
\bezier{180}(10.00,50.00)(36.00,34.00)(50.00,34.00)
\put(110.00,71.00){\makebox(0,0)[lb]{$\phi=$const}}
\put(70.00,29.00){\makebox(0,0)[cc]{$r=$ const}}
\put(50.50,65.00){\makebox(0,0)[lb]{$t=$ const}}
\put(82.00,29.00){\vector(2,1){16.00}}
\put(59.00,28.00){\vector(-3,-1){15.00}}
\put(22.00,62.00){\makebox(0,0)[cc]{$r$=0}}
\end{picture}
\caption{Conformal diagram of the ``extremal" Schwarzschild coordinates 
of Eq (\ref{horo}) in sections of AdS space. 
({\bf a}) An $r,\,t$ section. 
({\bf b}) An $r,\,\phi$ section. The lines $r=$ const are {\it horocycles} of the
Poincar\'e disk.}
\label{fig3} 
\end{figure}
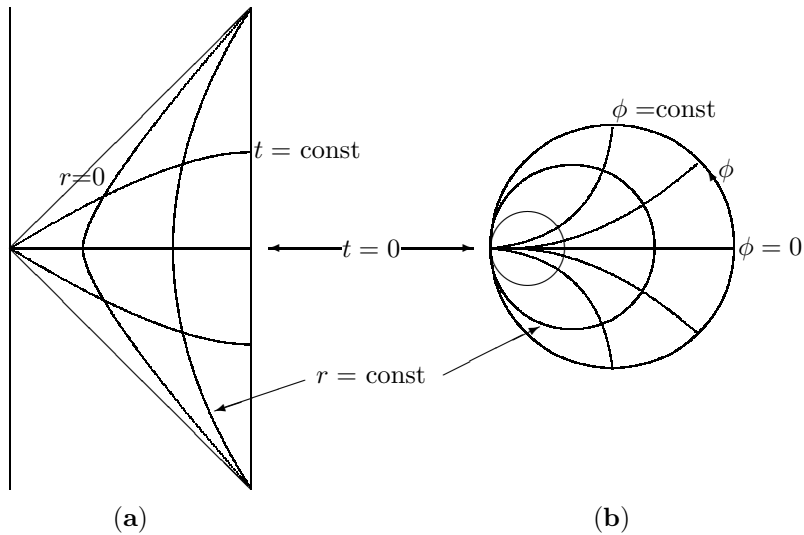  
 
The spacelike surfaces $t=$ const are conformally flat as are all two-dimensional surfaces, 
and as is manifest in Eq (\ref{sausge}). Less trivially, the three-dimensional AdS spacetime 
also has this property, so neighborhoods of AdS space can be conformally mapped to flat
space (one of the few cases where a three-dimensional conformal diagram 
exists). Such a map is the ``stereographic" projection, a projection by 
straight lines in the embedding space from a point in the surface of Eq (\ref{ads})
onto a plane tangent to that surface at the antipodal point, analogous to the 
familiar stereographic projection of a sphere (Fig.~4a).
By projection from the point $(U,\,V,\,X,\,Y) = (-\ell,\,0,\,0,\,0)$
to the plane $U=\ell$ we obtain the coordinates (provided $U > -\ell$)
\begin{equation}
	x^\mu = {2\ell X^\mu\over U+\ell} \qquad X^\mu \neq U
\label{stc}
\end{equation}
with the metric (where $X^0=V,\,x^0=t$)
\begin{equation}
	ds^2=\left({1\over 1-r_c^2}\right)^2\left(-dt^2+dx^2+dy^2\right)\quad
		{\rm where} \quad r_c^2={-t^2+x^2+y^2\over 4\ell^2}\,.
\label{stm}
\end{equation}
This metric is time-symmetric about $t=0$ but not static. It remains invariant 
under the Lorentz group of the flat 2+1-dimensional Minkowski space ($t,\,x,\,y$). 
In addition the origin may be shifted and the projection ``centered" about any 
point in AdS space (by projecting from the corresponding antipodal point).

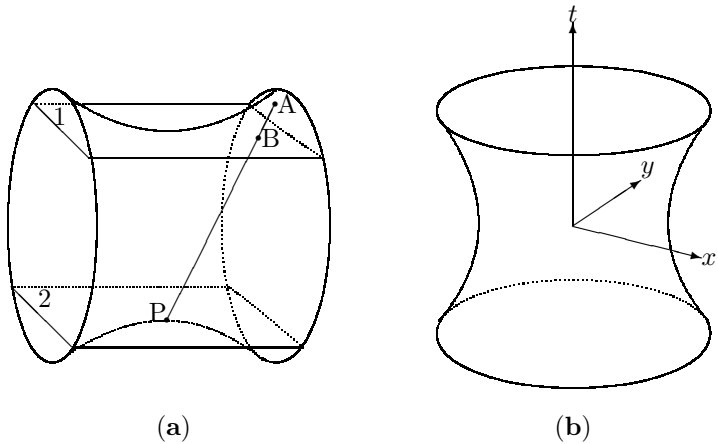
\begin{figure}
\unitlength 0.90mm
\linethickness{0.4pt}
\begin{picture}(127.20,65.00)(12,0)
\bezier{184}(63.00,55.20)(67.47,55.20)(70.00,45.10)
\bezier{180}(70.00,45.10)(71.99,35.00)(70.00,24.90)
\bezier{184}(70.00,24.90)(67.47,14.80)(63.00,14.80)
\bezier{30}(63.00,55.20)(58.53,55.20)(56.00,45.10)
\bezier{30}(56.00,45.10)(54.01,35.00)(56.00,24.90)
\bezier{30}(56.00,24.90)(58.53,14.80)(63.00,14.80)
\bezier{184}(30.00,55.20)(33.67,55.20)(35.76,45.10)
\bezier{180}(35.76,45.10)(37.40,35.00)(35.76,24.90)
\bezier{184}(35.76,24.90)(33.67,14.80)(30.00,14.80)
\bezier{184}(30.00,55.20)(26.33,55.20)(24.24,45.10)
\bezier{180}(24.24,45.10)(22.60,35.00)(24.24,24.90)
\bezier{184}(24.24,24.90)(26.33,14.80)(30.00,14.80)
\bezier{160}(31.00,55.00)(47.00,43.00)(63.00,55.00)
\bezier{24}(63.00,55.20)(61.51,55.20)(59.00,53.00)
\put(33.00,17.00){\line(-1,1){9.00}}   
\put(70.00,45.00){\line(-1,0){34.50}}
\put(35.50,45.00){\line(-1,1){8.20}}
\put(59.00,53.00){\line(-1,0){24.50}}
\bezier{28}(70.00,45.00)(63.67,49.07)(59.00,53.00)   
\bezier{14}(34.00,53.00)(31.63,52.96)(27.00,53.00)  
\put(33.00,17.00){\line(1,0){34.00}}
\bezier{28}(67.00,17.00)(60.70,22.59)(56.00,26.00)
\bezier{64}(56.00,26.00)(39.96,25.93)(24.00,26.00)
\put(47.00,21.00){\circle*{0.80}}
\bezier{80}(61.00,15.20)(47.00,26.85)(31.00,15.00) 
\put(47.00,21.00){\line(1,2){16.00}}
\put(63.30,53.00){\makebox(0,0)[lc]{A}}
\put(63.00,53.00){\circle*{0.80}}
\put(60.50,48.00){\circle*{0.80}}
\put(61.00,48.00){\makebox(0,0)[lc]{B}}
\put(47.00,20.93){\makebox(0,0)[rb]{P}}
\put(28.00,23.00){\makebox(0,0)[lb]{2}}
\put(30.20,51.20){\makebox(0,0)[lc]{1}}
\bezier{184}(126.80,20.90)(129.00,14.93)(117.10,12.00)
\bezier{180}(117.10,12.00)(107.00,10.01)(96.90,12.00)
\bezier{184}(96.90,12.00)(86.80,14.53)(86.80,19.00)
\bezier{30}(127.20,19.00)(127.20,23.47)(117.10,26.00)
\bezier{30}(117.10,26.00)(107.00,27.99)(96.90,26.00)
\bezier{30}(96.90,26.00)(86.80,23.47)(86.80,19.00)
\bezier{184}(127.20,52.00)(127.20,48.33)(117.10,46.24)
\bezier{184}(127.20,52.00)(127.20,48.33)(117.10,46.24)
\bezier{180}(117.10,46.24)(107.00,44.60)(96.90,46.24)
\bezier{184}(96.90,46.24)(86.80,48.33)(86.80,52.00)
\bezier{184}(127.20,52.00)(127.20,55.67)(117.10,57.76)
\bezier{180}(117.10,57.76)(107.00,59.40)(96.90,57.76)
\bezier{184}(96.90,57.76)(86.80,55.67)(86.80,52.00)
\bezier{160}(127.00,51.00)(115.00,35.00)(126.80,21.00)
\bezier{180}(87.20,21.00)(98.85,35.00)(87.00,51.00)
\put(48.00,5.00){\makebox(0,0)[cc]{({\bf a})}}
\put(107.00,5.00){\makebox(0,0)[cc]{({\bf b})}}
\put(107.00,35.00){\vector(0,1){30.00}}
\put(107.00,35.00){\vector(4,-1){19.00}}
\put(107.00,35.00){\vector(3,2){10.00}}
\put(117.00,42.00){\makebox(0,0)[lb]{$y$}}
\put(126.00,30.00){\makebox(0,0)[lc]{$x$}}
\put(107.00,65.20){\makebox(0,0)[cb]{$t$}}
\end{picture}
\caption{AdS space in stereographic projection.
({\bf a}) The hyperboloid is 2-dimensional AdS space embedded in 3-dimensional flat 
space as in Eq (\ref{ads}), restricted to $Y=0$. It is projected from 
point P onto the plane 1 ($U=\ell$). The image of 
point A in the hyperboloid is point B in the plane. The part of the hyperboloid that 
lies below plane 2 is not covered by the stereographic coordinates. 
({\bf b}) When plotted in the stereographic coordinates (\ref{stc}), AdS space is the 
interior of a hyperboloid. The boundary of the hyperboloid is (part of) the conformal 
boundary of AdS space.} \label{fig4} 
\end{figure}  

Because of the condition $U>-\ell$ the stereographic projection fails to cover a part of 
AdS space, even in the periodically identified version (Fig.~4a). The 3-dimensional conformal 
diagram is the interior of the hyperboloid $r_c=1$, where the conformal factor of the 
metric (\ref{stm}) is finite (Fig.~4b). On the surface of time-symmetry, $t=0$, the 
stereographic metric agrees with the sausage metric (\ref{sausge}).

Many similar coordinate systems, 
illustrating various symmetries of AdS space, are possible; for examples
see \cite{Li}.

\subsection{Isometries and Geodesics} 

To discuss the identifications that lead to time-symmetric black holes and 
other globally non-trivial 2+1-dimensional solutions we need a convenient 
representation of isometries and other geometrical relations in a spacelike
initial surface of time-symmetry. Such a representation is the conformal 
map of Figs 1 and 2, in which this spacelike surface is shown as a disk, 
known as the {\it Poincar\'e disk}. This representation has been 
extensively studied (see, for example, \cite{BP}), and we only mention the 
features that are most important for our task.

\begin{figure}
\unitlength 0.63mm
\linethickness{0.4pt}
\begin{picture}(144.00,109.00)
\bezier{436}(43.00,95.00)(80.00,55.00)(117.00,95.00)
\bezier{160}(43.00,96.00)(42.67,100.04)(60.00,102.83)
\bezier{188}(60.00,102.83)(80.00,105.48)(100.00,102.83)
\bezier{160}(100.00,102.83)(115.67,100.92)(117.00,96.00)
\bezier{160}(43.00,96.00)(42.67,91.96)(60.00,89.17)
\bezier{190}(60.00,89.17)(80.00,86.52)(100.00,89.17)
\bezier{160}(100.00,89.17)(115.67,91.08)(117.00,96.00)
\bezier{100}(43.00,15.00)(80.00,55.00)(117.00,15.00)
\bezier{160}(41.15,75.00)(40.80,79.24)(59.00,82.17)
\bezier{188}(59.00,82.17)(80.00,84.95)(101.00,82.17)
\bezier{160}(101.00,82.17)(117.45,80.17)(118.85,75.00)
\bezier{160}(41.15,75.00)(40.80,70.76)(59.00,67.83)
\bezier{188}(59.00,67.83)(80.00,65.05)(101.00,67.83)
\bezier{160}(101.00,67.83)(117.45,69.83)(118.85,75.00)
\bezier{70}(40.00,95.00)(60.00,75.00)(80.00,55.00)
\bezier{70}(80.00,55.00)(98.00,73.00)(120.00,95.00)
\bezier{40}(43.00,14.00)(42.67,18.04)(60.00,20.83)
\bezier{45}(60.00,20.83)(80.00,23.48)(100.00,20.83)
\bezier{40}(100.00,20.83)(115.67,18.92)(117.00,14.00)
\bezier{40}(43.00,14.00)(42.67,9.96)(60.00,7.17)
\bezier{45}(60.00,7.17)(80.00,4.52)(100.00,7.17)
\bezier{40}(100.00,7.17)(115.67,9.08)(117.00,14.00)
\put(122.00,109.00){\vector(-1,-1){9.00}}
\put(124.00,109.00){\makebox(0,0)[lc]{hyperboloid}}   
\put(142.00,75.00){\vector(-1,0){10.00}}
\put(144.00,75.00){\makebox(0,0)[lc]{plane}}
\put(28.00,69.00){\makebox(0,0)[rc]{limit circle}}  
\put(125.00,7.00){\vector(-4,3){8.00}}
\put(126.00,8.00){\makebox(0,0)[lc]{other sheet of hyperboloid}}
\put(126.00,5.50){\makebox(0,0)[lt]{(not used in construction)}}
\put(29.00,69.00){\vector(1,0){23.00}}
\put(123.00,85.00){\line(5,-6){16.67}}
\thicklines
\put(140.00,65.00){\line(-1,0){105.00}}
\put(35.00,65.00){\line(-3,5){12.00}}
\thinlines
\put(23.00,85.00){\line(1,0){100.00}}
\put(80.00,55.00){\vector(0,1){40.00}}
\put(80.00,96.00){\makebox(0,0)[cb]{$U$}}
\put(80.00,55.00){\vector(4,-1){30.00}}  
\put(111.00,47.00){\makebox(0,0)[lc]{$X$}}
\put(80.00,55.00){\vector(3,2){12.00}}  
\put(93.00,63.00){\makebox(0,0)[lc]{$Y$}}   
\put(80.00,45.00){\makebox(0,0)[cc]{$\ell$}}
\put(80.00,48.00){\vector(0,1){7.00}} 
\put(80.00,43.00){\vector(0,-1){8.00}}
\bezier{50}(118.00,73.00)(99.00,54.00)(80.00,35.00)
\bezier{50}(80.00,35.00)(60.67,54.33)(43.00,72.00)
\put(80.00,35.00){\line(-1,2){27.00}} 
\put(53.00,89.00){\circle*{1.0}}
\put(61.00,73.00){\circle*{1.0}}
\put(80.00,35.00){\circle*{1.0}}
\put(54.00,89.00){\makebox(0,0)[lc]{A}}
\put(62.00,73.00){\makebox(0,0)[lc]{B}}
\put(80.00,34.00){\makebox(0,0)[ct]{P}}
\end{picture}
\caption{The two-dimensional space $H^2$ of constant curvature $1/\ell^2$ is 
embedded in flat Minkowski space as one sheet of the hyperboloid of Eq 
(\ref{emb}). Under a stereographic projection from point P to the plane, 
point A on the hyperboloid is mapped to point B in the plane. Thus the 
hyperboloid ($H^2$) is mapped onto the Poincar\'e disk, the interior of 
the curve marked ``limit circle"}
\end{figure}
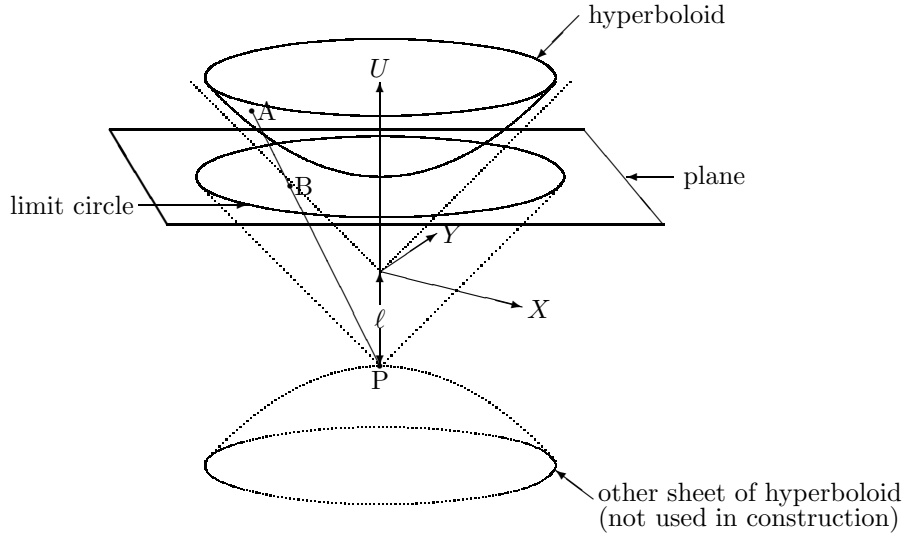  

All totally geodesic, time-symmetric surfaces $H^2$ in AdS space are 
isometric to the typical hyperboloid (Fig.~5) obtained by restricting 
Eq (\ref{ads}) to $V=0$,
\begin{equation}
	X^2+Y^2-U^2=-\ell^2     \label{emb}
\end{equation}
This surface has zero extrinsic curvature and therefore constant negative 
Gaussian curvature $-1/\ell^2$. The Poincar\'e 
disk can be obtained as a map of $H^2$ by the stereographic projection of 
Fig.~5, which illustrates Eq (\ref{stc}) when restricted to $V=0$ 
similar to the way Fig.~4 illustrates it when restricted to $X=0$. In 
this way 
all of $H^2$ is mapped into the interior of a disk of radius $2\ell$, 
whose boundary, called the limit circle, represents points at 
(projective or conformal) infinity. Because the map is conformal,
angles are faithfully represented. Other geometrical objects in $H^2$ appear
distorted in the Euclidean geometry of the disk, but by assigning new 
roles to these ``distorted" objects and manipulating those according to 
Euclidean geometry one can perform constructions equivalent to those 
in the $H^2$-geometry directly on the Poincar\'e disk. 

For example, on the surface $H^2$ as described by Eq (\ref{emb}), 
all geodesics are intersections of planes through the origin with the
surface; that is, they satisfy a linear relation between 
$X,\,Y,\,U$. From Eq (\ref{stc}) it follows directly that Eq (\ref{emb})
becomes such a linear relation if $x,\,y$ satisfy the equation of a circle that has
radius $(a^2-4\ell^2)$ if it is centered at $(x,\,y)=(a_x,\,a_y)$, hence meets
the limit circle at right angles. Because two such circles intersect in 
at most one point in the interior of the Poincar\'e disk, it follows that 
two geodesics in $H^2$ meet at most in one point (as in Euclidean space).

An important difference occurs if two geodesics do {\it not} meet:
in Euclidean space they are then equidistant; whereas in the Poincar\'e
disc the geodesic between points on two disjoint geodesics (Euclidean circles 
perpendicular to the limit circle) approaches a complete geodesic 
as the points approach the limit circle. Since 
the conformal factor in the metric of Eq (\ref{stm}), 
restricted to $t=0$,
\begin{equation}
ds^2 = \left(1-{x^2+y^2\over 4\ell^2}\right)^{-2}\left(dx^2+dy^2\right)
	\label{cm}
\end{equation}
increases without limit as $x^2+y^2\rightarrow 4\ell^2$,
on $H^2$ the geodesic distance between two given disjoint geodesics typically 
increases without bound as we go along the given geodesics in either 
direction. However, the geodesic distance between points on two given disjoint
geodesics of course has a lower bound.
If this is nonzero there is a unique geodesic segment of minimal 
length joining the two given geodesics at right angles to either.

On the other hand, if we have a family of equidistant curves, at most one of 
them can be a geodesic, and then the representation of the others on the
Poincar\'e disk are arcs of circles, not perpendicular to the limit circle, 
but meeting the geodesic asymptotically at the limit circle. 
The curves $r=$ const of Fig.~2b are 
examples, with $r=\ell$ the geodesic of the family. These equidistant curves 
have constant acceleration (with respect to their arclength parameter), and they 
also illustrate how the conformal factor in (\ref{cm}) distorts the apparent 
(Euclidean) distances of the disk into the true distances of $H^2$.

Because the surface (\ref{emb}) in Minkowski space has constant extrinsic
curvature, any isometry of the surface geometry can be extended to an
isometry of the embedding space. But we know all those isometries: they
form the homogeneous isochronous Lorentz group. Thus any Lorentz transformation
implies, by the projection of Fig.~5, a corresponding transformation of the
Poincar\'e disk that represents an isometry of $H^2$, and all $H^2$
isometries can be obtained in this way. In the Euclidean metric of the
disk such transformations must be conformal transformations leaving the
limit circle fixed, since they
are isometries of the conformal metric (\ref{cm}).

Knowing this we can now classify\footnote{We confine
attention to orientation-preserving transformations; they can be combined with
a reflection about a geodesic (with an infinite number of fixed
points) to obtain the rest.} the isometries of $H^2$. Proper 
Lorentz transformations in 3D Minkowski space have an axis of fixed points 
that may be a spacelike, null, or timelike straight line. If the axis is
timelike, it intersects the hyperboloid (\ref{emb}). If the axis is null,
it intersects the hyperboloid asymptotically. If the axis is spacelike,
it does not intersect the hyperboloid, but there are two fixed
null directions perpendicular to the axis. Correspondingly on the Poincar\'e
disk there is either one fixed point within the disk (``elliptic"), or one fixed point
on the limit circle (``parabolic"), or two fixed points on the limit circle 
(``hyperbolic") for these transformations. Fig.~1b
illustrates by the transformation $\theta \rightarrow \theta+$const the case 
with one finite fixed point 
(the origin). Figs.~2b and 3b illustrate by the transformation $\phi \rightarrow \phi+$const 
the case with two fixed points and one fixed point, respectively, on the limit circle ($\phi 
=\pm\infty$). In the case of two fixed points there is a unique geodesic 
($r=\ell$ in Fig.~2b) 
left fixed by the isometry, and conversely the isometry, which we will call ``along" the 
geodesic, is uniquely defined by the invariant geodesic and the distance by which a point 
moves along that geodesic.

Except for the rotation about the center of the disk as in Fig.~1b these
are not isometries of the disk's flat, Euclidean metric, but they are 
of course conformal isometries of this metric. Such conformal 
transformations, mapping the limit
circle into itself, are conveniently described as {\em M\"obius 
transformations} of the complex coordinate
\begin{equation}
	z = {x+iy\over\ell} \quad {\rm by} \quad z\rightarrow z'={az+b\over \bar b z+\bar a}\,,
\label{moeb}
\end{equation}
where $a,\,b$ are complex numbers with $|a|^2-|b|^2=1$. When we consider an
isometry or identification abstractly, it can always be implemented concretely
by such a M\"obius transformation. In particular, hyperbolic isometries
are described by M\"obius transformations with real $a$.

As the examples of Figs.~1-3 show, each of these isometries is part of a family
depending on a continuous parameter (the constant in $\phi\rightarrow\phi
+$const, for example). There is therefore an ``infinitesimal" version of
each isometry, described by a Killing vector ($\partial/\partial\phi$ in the
example). Conversely an (orientation-preserving) isometry can be described as the
exponential of its Killing vector.

\subsection{Identifications} 

The hyperbolic transformations, which have no fixed points in $H^2$,
are suitable for forming nonsingular quotient spaces that have the same local geometry
as AdS space, and hence satisfy the same Einstein equations. In the 
context of Fig.~2b
and Eq (\ref{schwmet}) the transformation that comes to mind is described by
$\phi\rightarrow\phi+2\pi$. The quotient space is the space in which points connected
by this transformation are regarded as identical, which is the same as the space
in which $\phi$ is a periodic coordinate with the usual period. Eq (\ref{schwmet}) with
this periodicity in $\phi$ already gives us the simplest BTZ metric for a single, non-rotating 
2+1-dimensional black hole. It is asymptotically AdS, as shown by comparing Eqs
(\ref{schwmet}) and (\ref{m0}).

The minimum distance between the two identified geodesics occurs at $r=\ell$ and is $2\pi\ell$. 
This is the minimum distance around the black hole, and plays the role of the 
horizon ``area". If we identify $\phi$ with a different 
period $2\pi a$, we get a metric with a different horizon size. We can then redefine the 
coordinates so that $\phi$ has its usual period,
$$\phi\rightarrow a\phi,\qquad r\rightarrow r/a\qquad t\rightarrow at$$
and the metric takes this standard form, called the BTZ metric \cite{BTZ}:
\begin{equation}
	ds^2=-\left({r^2\over\ell^2}-m\right) dt^2+\left({r^2\over\ell^2}-m\right)^{-1} dr^2+
	r^2 d\phi^2,
	\label{BTZ}
\end{equation}
where $m=1/a^2$. Here the dimensionless quantity $m$ is called the mass parameter. Although it can 
be measured in the asymptotic region, it is more directly related to the 
horizon size, the length of the minimal geodesic at the horizon, 
$2\pi\ell\sqrt{m}$.

The metric (\ref{BTZ}) is a solution also for $m=0$, as shown by Eq (\ref{horom}), but that is not the 
AdS metric itself. The latter is also described by Eq (\ref{BTZ}), but with $m=-1$, as shown by 
Eq (\ref{m0}). By contrast, the $m=0$ initial state is obtained by identifying the 
geodesics $\phi=0$ and $\phi=2\pi$ in Fig.~3b.

To describe the identification more explicitly, we may say that we have cut a strip from $\phi=0$ to 
$\phi=2\pi$ out of Fig.~2b, and glued the edges together. This strip is a ``fundamental domain" for
our identification, a region that contains images of its own points under the group only
on its boundary, and that together with all its images covers the full AdS space. To
obtain a fundamental domain for the BTZ black hole we might have used as the boundaries
some other curve on the Poincar\'e disk and its image under the transformation, provided only 
that the curve and its image do not intersect. But since it is always possible to avoid 
apparent asymmetries by choosing boundaries composed of geodesics that meet at right angles, we will 
generally do so.

We can think of the identification in yet another way, by a process that has
been called ``doubling": 
cut a strip from $\phi=0$ to $\phi=\pi$ from Fig.~2b, and cut another identical strip.
Put one on top of the other and glue the two edges together, obtaining 
again the black hole initial state. The gluing makes the two strips reflections of each other
with respect to either of the original edges. Back on the Poincar\'e disk the composition of
the two reflections is a translation in $\phi$ by $2\pi$, that is, the isometry of the identification.
Any (orientation-preserving) isometry of a hyperbolic space can be decomposed into two reflections
\cite{Bach};
hence any quotient space can be considered the double of a suitable region (possibly in another
quotient space), and a fundamental domain is obtained from the region and one of its reflections. 
 
The process of gluing together a constant negative curvature space from a fundamental 
domain of the Poincar\'e disk can be reversed: we cut the space by geodesics into
its fundamental domain, make many copies of the domain, and put these down on the disk so that
boundaries coming from the same cut touch, until the entire disk is covered. The resulting pattern is 
called a ``tiling" of the disk (although the ``tiles" corresponding to the $t=0$ section of the BTZ 
black hole look more like strip flooring). Thus we have two equivalent ways of describing our 
identified space: by giving a fundamental domain and rules of gluing the boundaries, or by a tiling 
together with rules relating each tile to its neighbors.

\begin{figure}
\unitlength 0.70mm
\linethickness{0.4pt}
\begin{picture}(150.00,82.00)(-5,0)
\thicklines
\bezier{40}(5.00,17.00)(5.00,47.00)(5.00,77.00)
\bezier{40}(5.00,77.00)(35.00,77.00)(65.00,77.00)
\bezier{40}(35.00,17.00)(35.00,47.00)(35.00,77.00)
\bezier{40}(65.00,17.00)(65.00,47.00)(65.00,77.00)
\bezier{40}(5.00,47.00)(35.00,47.00)(65.00,47.00)
\bezier{40}(5.00,17.00)(35.00,17.00)(65.00,17.00)
\put(35.00,47.00){\line(5,3){9.00}}
\put(65.00,47.00){\line(-3,5){21.00}}
\put(44.00,82.00){\line(-5,-3){22.00}}
\put(65.00,47.00){\line(-5,-3){13.00}}
\put(52.00,39.20){\line(-3,5){7.68}}
\put(65.00,47.00){\circle*{2.00}}
\put(35.00,47.00){\circle*{2.00}}
\put(57.00,60.00){\circle*{2.00}}
\put(35.00,47.00){\line(-3,5){13.20}}
\thinlines
\put(22.00,39.00){\line(5,3){36.00}}
\put(27.20,60.00){\line(-5,-3){22.20}}
\put(27.20,30.00){\line(-5,-3){22.20}}
\put(57.20,30.00){\line(-5,-3){22.20}}
\put(65.00,17.00){\line(-5,-3){13.00}}
\put(52.00,9.20){\line(-3,5){7.68}}
\put(65.00,77.00){\line(-5,-3){13.00}}
\put(35.00,47.00){\line(-3,5){21.00}}
\put(14.00,82.00){\line(-5,-3){9.00}}
\put(35.00,17.00){\line(-3,5){21.00}}
\put(65.00,17.00){\line(-3,5){13.80}}
\put(35.00,17.00){\line(-5,-3){13.00}}
\put(22.00,9.20){\line(-3,5){7.68}}
\thicklines
\put(108.00,40.00){\line(0,1){12.00}}
\put(108.00,52.00){\line(1,0){12.00}}
\put(120.00,52.00){\line(0,1){18.00}}
\put(108.00,40.00){\line(-1,0){18.00}}
\put(90.00,40.00){\line(0,1){30.00}}
\put(90.00,70.00){\line(1,0){30.00}}
\bezier{23}(90.00,40.00)(105.00,46.00)(120.00,52.00)
\bezier{23}(90.00,40.00)(96.00,55.00)(102.00,70.00)
\bezier{23}(108.00,22.00)(123.00,28.00)(138.00,34.00)
\bezier{23}(108.00,22.00)(114.00,37.00)(120.00,52.00)
\bezier{23}(120.00,52.00)(135.00,58.00)(150.00,64.00)
\bezier{23}(120.00,52.00)(126.00,67.00)(132.00,82.00)
\bezier{23}(138.00,34.00)(144.00,49.00)(150.00,64.00)
\bezier{23}(78.00,10.00)(84.00,25.00)(90.00,40.00)
\bezier{23}(102.00,70.00)(117.00,76.00)(132.00,82.00)
\bezier{23}(78.00,10.00)(93.00,16.00)(108.00,22.00)
\thinlines
\put(138.00,34.00){\line(0,1){18.00}}
\put(126.00,22.00){\line(-1,0){18.00}}
\put(120.00,52.00){\line(1,0){18.00}}
\put(108.00,40.00){\line(0,-1){18.00}}
\put(108.00,40.00){\line(1,0){12.00}}
\put(120.00,40.00){\line(0,1){12.00}}
\put(120.00,52.00){\circle*{2.00}}
\put(120.00,40.00){\circle*{2.00}}
\put(90.00,40.00){\circle*{2.00}}
\put(96.00,10.00){\line(0,1){12.00}}
\put(96.00,22.00){\line(1,0){12.00}}
\put(96.00,10.00){\line(-1,0){18.00}}
\put(78.00,10.00){\line(0,1){30.00}}
\put(78.00,40.00){\line(1,0){12.00}}
\put(120.00,70.00){\line(0,1){12.00}}
\put(138.00,52.00){\line(0,1){12.00}}
\put(138.00,64.00){\line(1,0){12.00}}
\put(150.00,64.00){\line(0,1){18.00}}
\put(120.00,82.00){\line(1,0){30.00}}
\put(138.00,34.00){\line(1,0){12.00}}
\put(126.00,22.00){\line(0,-1){12.00}}
\put(102.00,70.00){\line(0,1){12.00}}
\put(90.00,58.00){\line(-1,0){12.00}}
\put(126.00,22.00){\line(0,1){12.00}}
\put(126.00,34.00){\line(1,0){12.00}}
\put(144.00,10.00){\line(0,1){6.00}}
\put(144.00,16.00){\line(1,0){6.00}}
\put(35.00,2.00){\makebox(0,0)[cc]{({\bf a})}}
\put(114.00,2.00){\makebox(0,0)[cc]{({\bf b})}}
\put(40.00,64.00){\makebox(0,0)[cc]{$a^2$}}
\put(54.00,50.00){\makebox(0,0)[cc]{$b^2$}}
\put(105.00,55.00){\makebox(0,0)[cc]{$a^2$}}
\put(114.00,46.00){\makebox(0,0)[cc]{$b^2$}}
\end{picture}
\caption{Two different ways of tiling the plane prove the theorem of
Pythagoras in ({\bf a}) Euclidean space and ({\bf b}) Minkowski space}
\end{figure}
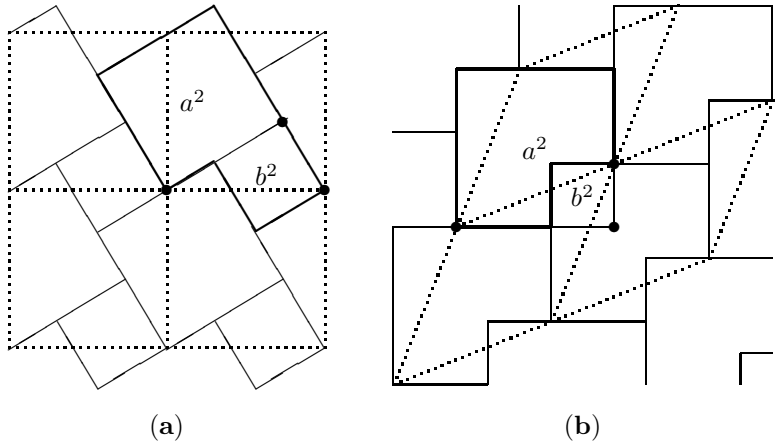

\subsubsection{Tiling and Pythagoras}

To fix ideas, consider an application of tiling found among the numerous
proofs of the theorem of Pythgoras (a local boy who contributed to the
early fame of Samos). This proof is based on the fact that all fundamental
domains of a given group of isometries have equal area. In the Euclidean
plane we consider the group generated by two translations specified in direction 
and amounts by two adjacent sides of the square above the hypotenuse of
a right triangle, whose vertices are the three larger dots in Fig.~6a. This
square is a fundamental domain of the group, and part of the tiling by
this square is shown by the 
horizontal and vertical dotted lines. The region drawn in heavy outline is 
an alternative fundamental domain of the same group of isometries, 
and that domain is made from the squares above the sides of the same triangle.
Part of its tiling of the plane is shown by the lightly drawn lines of
Fig.~6a.
Either fundamental domain can be glued together to form the same quotient 
space, a ``square" torus, so the areas are equal, $c^2 = a^2+b^2 =$ area of
torus.

In special relativity the theorem of Pythagoras is valid with a different
sign, $c^2=a^2-b^2$ if we choose the hypotenuse and one of the sides
to be spacelike, and of course the right angles of the triangle and of squares 
are to be drawn in accordance with the Minkowski metric. Fig.~6b shows the
proof by the tiling that derives from a Minkowski torus of area $c^2$.
(Here we use, at least implicitly, the fact that the area of a two-dimensional
figure is the same in Euclidean and Minkowski spaces if their metrics 
differ only by a sign.)

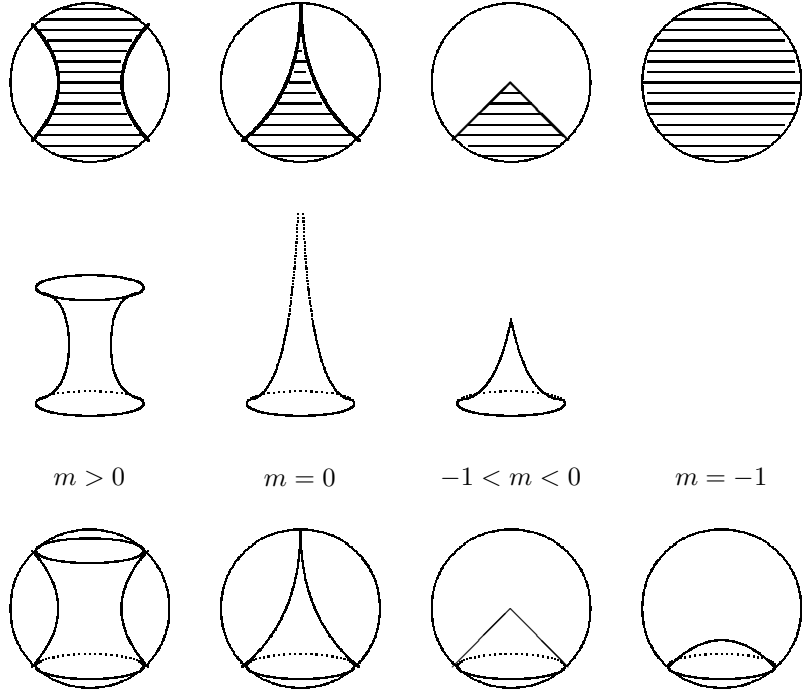
\begin{figure}
\unitlength 0.70mm
\linethickness{0.4pt}
\begin{picture}(155.08,136.96)
\bezier{50}(4.92,20.00)(4.92,28.41)(12.46,33.19)
\bezier{50}(12.46,33.19)(20.00,36.96)(27.54,33.19)
\bezier{50}(27.54,33.19)(35.08,28.41)(35.08,20.00)
\bezier{50}(4.92,20.00)(4.92,11.59)(12.46,6.81)
\bezier{50}(12.46,6.81)(20.00,3.04)(27.54,6.81)
\bezier{50}(27.54,6.81)(35.08,11.59)(35.08,20.00)
\bezier{50}(44.92,20.00)(44.92,28.41)(52.46,33.19)
\bezier{50}(84.92,20.00)(84.92,28.41)(92.46,33.19)
\bezier{50}(124.92,20.00)(124.92,28.41)(132.46,33.19)
\bezier{50}(52.46,33.19)(60.00,36.96)(67.54,33.19)
\bezier{50}(92.46,33.19)(100.00,36.96)(107.54,33.19)
\bezier{50}(132.46,33.19)(140.00,36.96)(147.54,33.19)
\bezier{50}(67.54,33.19)(75.08,28.41)(75.08,20.00)
\bezier{50}(107.54,33.19)(115.08,28.41)(115.08,20.00)
\bezier{50}(147.54,33.19)(155.08,28.41)(155.08,20.00)
\bezier{50}(44.92,20.00)(44.92,11.59)(52.46,6.81)
\bezier{50}(84.92,20.00)(84.92,11.59)(92.46,6.81)
\bezier{50}(124.92,20.00)(124.92,11.59)(132.46,6.81)
\bezier{50}(52.46,6.81)(60.00,3.04)(67.54,6.81)
\bezier{50}(92.46,6.81)(100.00,3.04)(107.54,6.81)
\bezier{50}(132.46,6.81)(140.00,3.04)(147.54,6.81)
\bezier{50}(67.54,6.81)(75.08,11.59)(75.08,20.00)
\bezier{50}(107.54,6.81)(115.08,11.59)(115.08,20.00)
\bezier{50}(147.54,6.81)(155.08,11.59)(155.08,20.00)
\put(89.00,9.00){\line(1,1){11.00}}
\put(100.00,20.00){\line(1,-1){11.00}}
\bezier{120}(49.00,9.00)(60.00,18.00)(60.00,35.00)
\bezier{120}(71.00,9.00)(60.00,18.00)(60.00,35.00)
\bezier{120}(31.00,9.00)(21.00,20.00)(31.00,31.00)
\bezier{120}(9.00,9.00)(19.00,20.00)(9.00,31.00)
\bezier{50}(4.92,120.00)(4.92,128.41)(12.46,133.19)
\bezier{50}(12.46,133.19)(20.00,136.96)(27.54,133.19)
\bezier{50}(27.54,133.19)(35.08,128.41)(35.08,120.00)
\bezier{50}(4.92,120.00)(4.92,111.59)(12.46,106.81)
\bezier{50}(12.46,106.81)(20.00,103.04)(27.54,106.81)
\bezier{50}(27.54,106.81)(35.08,111.59)(35.08,120.00)
\bezier{50}(44.92,120.00)(44.92,128.41)(52.46,133.19)
\bezier{50}(84.92,120.00)(84.92,128.41)(92.46,133.19)
\bezier{50}(124.92,120.00)(124.92,128.41)(132.46,133.19)
\bezier{50}(52.46,133.19)(60.00,136.96)(67.54,133.19)
\bezier{50}(92.46,133.19)(100.00,136.96)(107.54,133.19)
\bezier{50}(132.46,133.19)(140.00,136.96)(147.54,133.19)
\bezier{50}(67.54,133.19)(75.08,128.41)(75.08,120.00)
\bezier{50}(107.54,133.19)(115.08,128.41)(115.08,120.00)
\bezier{50}(147.54,133.19)(155.08,128.41)(155.08,120.00)
\bezier{50}(44.92,120.00)(44.92,111.59)(52.46,106.81)
\bezier{50}(84.92,120.00)(84.92,111.59)(92.46,106.81)
\bezier{50}(124.92,120.00)(124.92,111.59)(132.46,106.81)
\bezier{50}(52.46,106.81)(60.00,103.04)(67.54,106.81)
\bezier{50}(92.46,106.81)(100.00,103.04)(107.54,106.81)
\bezier{50}(132.46,106.81)(140.00,103.04)(147.54,106.81)
\bezier{50}(67.54,106.81)(75.08,111.59)(75.08,120.00)
\bezier{50}(107.54,106.81)(115.08,111.59)(115.08,120.00)
\bezier{50}(147.54,106.81)(155.08,111.59)(155.08,120.00)
\thicklines
\put(89.00,109.00){\line(1,1){11.00}}
\put(100.00,120.00){\line(1,-1){11.00}} 
\bezier{120}(49.00,109.00)(60.00,118.00)(60.00,135.00)
\bezier{120}(71.00,109.00)(60.00,118.00)(60.00,135.00)
\bezier{120}(31.00,109.00)(21.00,120.00)(31.00,131.00)
\bezier{120}(9.00,109.00)(19.00,120.00)(9.00,131.00)
\thinlines
\bezier{10}(9.79,9.00)(9.79,10.32)(14.90,11.07)
\bezier{10}(14.90,11.07)(20.00,11.66)(25.10,11.07)
\bezier{10}(25.10,11.07)(30.21,10.32)(30.21,9.00)
\bezier{50}(9.79,9.00)(9.79,7.68)(14.90,6.93)
\bezier{50}(14.90,6.93)(20.00,6.34)(25.10,6.93)
\bezier{50}(25.10,6.93)(30.21,7.68)(30.21,9.00)
\bezier{50}(9.79,31.00)(9.79,32.32)(14.90,33.07)
\bezier{50}(14.90,33.07)(20.00,33.66)(25.10,33.07)
\bezier{50}(25.10,33.07)(30.21,32.32)(30.21,31.00)
\bezier{50}(9.79,31.00)(9.79,29.68)(14.90,28.93)
\bezier{50}(14.90,28.93)(20.00,28.34)(25.10,28.93)
\bezier{50}(25.10,28.93)(30.21,29.68)(30.21,31.00)
\bezier{10}(49.79,9.00)(49.79,10.32)(54.90,11.07)
\bezier{10}(54.90,11.07)(60.00,11.66)(65.10,11.07)
\bezier{10}(65.10,11.07)(70.21,10.32)(70.21,9.00)
\bezier{50}(49.79,9.00)(49.79,7.68)(54.90,6.93)
\bezier{50}(54.90,6.93)(60.00,6.34)(65.10,6.93)
\bezier{50}(65.10,6.93)(70.21,7.68)(70.21,9.00)
\bezier{10}(89.79,9.00)(89.79,10.32)(94.90,11.07)
\bezier{10}(94.90,11.07)(100.00,11.66)(105.10,11.07)
\bezier{10}(105.10,11.07)(110.21,10.32)(110.21,9.00)
\bezier{50}(89.79,9.00)(89.79,7.68)(94.90,6.93)
\bezier{50}(94.90,6.93)(100.00,6.34)(105.10,6.93)
\bezier{50}(105.10,6.93)(110.21,7.68)(110.21,9.00)
\bezier{10}(129.79,9.00)(129.79,10.32)(134.90,11.07)
\bezier{10}(134.90,11.07)(140.00,11.66)(145.10,11.07)
\bezier{10}(145.10,11.07)(150.21,10.32)(150.21,9.00)
\bezier{50}(129.79,9.00)(129.79,7.68)(134.90,6.93)
\bezier{50}(134.90,6.93)(140.00,6.34)(145.10,6.93)
\bezier{50}(145.10,6.93)(150.21,7.68)(150.21,9.00)
\bezier{112}(130.00,9.00)(140.00,19.00)(150.00,9.00)
\bezier{10}(9.79,59.00)(9.79,60.32)(14.90,61.07)
\bezier{10}(14.90,61.07)(20.00,61.66)(25.10,61.07)
\bezier{10}(25.10,61.07)(30.21,60.32)(30.21,59.00)
\bezier{50}(9.79,59.00)(9.79,57.68)(14.90,56.93)
\bezier{50}(14.90,56.93)(20.00,56.34)(25.10,56.93)
\bezier{50}(25.10,56.93)(30.21,57.68)(30.21,59.00)
\bezier{50}(9.79,81.00)(9.79,82.32)(14.90,83.07)
\bezier{50}(14.90,83.07)(20.00,83.66)(25.10,83.07)
\bezier{50}(25.10,83.07)(30.21,82.32)(30.21,81.00)
\bezier{50}(9.79,81.00)(9.79,79.68)(14.90,78.93)
\bezier{50}(14.90,78.93)(20.00,78.34)(25.10,78.93)
\bezier{50}(25.10,78.93)(30.21,79.68)(30.21,81.00)
\bezier{10}(49.79,59.00)(49.79,60.32)(54.90,61.07)
\bezier{10}(89.79,59.00)(89.79,60.32)(94.90,61.07)
\bezier{10}(54.90,61.07)(60.00,61.66)(65.10,61.07)
\bezier{10}(94.90,61.07)(100.00,61.66)(105.10,61.07)
\bezier{10}(65.10,61.07)(70.21,60.32)(70.21,59.00)
\bezier{10}(105.10,61.07)(110.21,60.32)(110.21,59.00)
\bezier{50}(49.79,59.00)(49.79,57.68)(54.90,56.93)
\bezier{50}(89.79,59.00)(89.79,57.68)(94.90,56.93)
\bezier{50}(54.90,56.93)(60.00,56.34)(65.10,56.93)
\bezier{50}(94.90,56.93)(100.00,56.34)(105.10,56.93)
\bezier{50}(65.10,56.93)(70.21,57.68)(70.21,59.00)
\bezier{50}(105.10,56.93)(110.21,57.68)(110.21,59.00)
\bezier{100}(50.00,59.70)(58.00,59.70)(59.50,95.00)
\bezier{88}(90.00,59.70)(97.00,59.70)(100.00,75.00)
\bezier{52}(11.00,80.00)(16.00,79.00)(16.00,70.00)
\bezier{64}(16.00,70.00)(16.00,61.00)(10.00,59.70)
\put(15.00,134.00){\line(1,0){10.00}}
\put(14.00,120.00){\line(1,0){12.00}}
\put(14.00,122.00){\line(1,0){12.00}}
\put(26.00,124.00){\line(-1,0){12.00}}
\put(13.00,126.00){\line(1,0){14.00}}
\put(28.00,128.00){\line(-1,0){16.00}}
\put(10.00,130.00){\line(1,0){20.00}}
\put(29.00,132.00){\line(-1,0){18.00}}
\bezier{50}(12.46,106.81)(20.00,103.04)(27.54,106.81)
\put(15.00,106.00){\line(1,0){10.00}}
\put(14.00,118.00){\line(1,0){12.00}}
\put(26.00,116.00){\line(-1,0){12.00}}
\put(13.00,114.00){\line(1,0){14.00}}
\put(28.00,112.00){\line(-1,0){16.00}}
\put(10.00,110.00){\line(1,0){20.00}}
\put(29.00,108.00){\line(-1,0){18.00}}
\put(55.00,106.00){\line(1,0){10.00}}
\put(70.00,110.00){\line(-1,0){20.00}}
\put(52.00,112.00){\line(1,0){16.00}}
\put(66.00,114.00){\line(-1,0){12.00}}
\put(56.00,116.00){\line(1,0){8.00}}
\put(57.00,118.00){\line(1,0){6.00}}
\put(62.00,120.00){\line(-1,0){4.00}}
\put(59.00,122.00){\line(1,0){2.00}}
\put(59.00,124.00){\line(1,0){2.00}}
\put(59.67,126.00){\line(1,0){0.67}}
\put(95.00,106.00){\line(1,0){11.00}}
\put(109.00,108.00){\line(-1,0){17.00}}
\put(90.00,110.00){\line(1,0){20.00}}
\put(108.00,112.00){\line(-1,0){16.00}}
\put(94.00,114.00){\line(1,0){12.00}}
\put(104.00,116.00){\line(-1,0){8.00}}
\put(98.00,118.00){\line(1,0){4.00}}
\put(135.00,134.00){\line(1,0){10.00}}
\put(131.00,132.00){\line(1,0){18.00}}
\put(129.00,130.00){\line(1,0){22.00}}
\put(128.00,128.00){\line(1,0){24.00}}
\put(154.00,126.00){\line(-1,0){28.00}}
\put(126.00,124.00){\line(1,0){28.00}}
\put(126.00,122.00){\line(1,0){28.00}}
\put(155.00,120.00){\line(-1,0){30.00}}
\put(129.00,110.00){\line(1,0){22.00}}
\put(128.00,112.00){\line(1,0){24.00}}
\put(154.00,114.00){\line(-1,0){28.00}}
\put(126.00,116.00){\line(1,0){28.00}}
\put(126.00,118.00){\line(1,0){28.00}}
\put(135.00,106.00){\line(1,0){10.00}}
\put(131.00,108.00){\line(1,0){18.00}}
\put(20.00,45.00){\makebox(0,0)[cc]{$m>0$}}
\put(60.00,45.00){\makebox(0,0)[cc]{$m=0$}}
\put(100.00,45.00){\makebox(0,0)[cc]{$-1<m<0$}}
\put(140.00,45.00){\makebox(0,0)[cc]{$m=-1$}}
\bezier{52}(29.00,80.00)(24.00,79.00)(24.00,70.00)
\bezier{64}(24.00,70.00)(24.00,61.00)(30.00,59.70)
\bezier{100}(70.00,59.70)(62.00,59.70)(60.50,95.00)
\bezier{88}(110.00,59.70)(103.00,59.70)(100.00,75.00)
\put(51.00,108.00){\line(1,0){18.00}}
\end{picture}
\caption{Three representations of the geometry of the $t=0$ geometry of metric 
(\ref{BTZ}) for different ranges of the mass parameter $m$: The BTZ
black hole for $m>0$; the extremal BTZ black hole for $m=0$; the
point particle (conical singularity) for $m<0$; and AdS (``vacuum")
space itself for $m=-1$.
{\bf Top row:} shaded regions of the Poincar\'e disk, to be identified 
in each figure along the left and right boundaries, drawn in thicker lines.
{\bf Second row:} an embedding of the central part ($r\leq\ell\sqrt{1-m}$)
of these spaces as surfaces in three-dimensional flat Euclidean space. 
The embedding cannot be continued beyond the outer edges of each figure.
{\bf Bottom row:} the entire surface can be embedded in a 3D space of constant 
negative curvature, shown as a Poincar\'e ball. (The figure is
schematic only; for example, the angle at the conical tips ought to be the
same in the second and last row, to represent the same surface)}\label{fig6}

\end{figure}

\subsubsection{Embeddings}

To visualize the geometry of our glued-together surface --- the $t=0$ surface of a static BTZ black 
hole --- it helps to embed this surface in a three-dimensional space in which the gluing can be 
actually carried out. This is analogous to the embedding of the $t=0$ surface of the Schwarzschild 
black hole, with one angle suppressed, in three-dimensional flat space (the surface of rotation of 
the {\it Flamm parabola} \cite{MTW}). For the BTZ initial surface only a finite part can 
be so embedded. The embedding stops where
the rate of increase of circumference of the circle $r=$ const with respect to the true distance 
in the radial direction exceeds that rate in flat space.
(The remainder of the surface could then be embedded in Minkowski space, but the switch between 
embeddings is an artifact and corresponds to no local intrinsic property.) However, the entire surface can be 
embedded in $H^3$, the {\it Riemannian} (positive definite metric) space of constant negative 
curvature. By the obvious generalization of the Poincar\'e disk this space can be conformally 
represented as a ball in three-dimensional flat space. Fig.~7 shows this embedding, where the
surface for $m>0$ is seen to have two asymptotic sheets, similar to the corresponding Schwarzschild surface.  

\subsection{Multiple Black Holes}

We saw that a single hyperbolic isometry (call it $a$) used as an identification to obtain an
AdS initial state always yields a (single) BTZ black hole state, with horizon
size and location depending on $a$. For other types of initial states we 
therefore need to use more
than one such isometry, for example $a$ and $b$. Assuring that there are no fixed points
(which would lead to singularities of the quotient space) would then seem to be much more 
difficult: If we know that $a$ has no fixed point, then the whole group consisting of powers
$a^n$ has no fixed points (except the identity, $n=0$); but for the group generated by two
isometries $a,\,b$ we have to check that no ``word" formed from these and their inverses,
such as $ab^{-2}a^3b$ has fixed points. Although this may seem complicated, it is easy if we
have a fundamental domain such that the isometry $a$ maps one of a pair of boundaries into the
other, and the isometry $b$ does the same for a different pair of boundaries. Now tile the
Poincar\'e disk with copies of this fundamental domain (see Fig.~10 for an example). 
Once we fix the original tile (associated
with the identity isometry), there is a one-to-one correspondence between tiles and words.
Therefore every non-trivial word moves all points in the original tile to some different tile,
and there can be no fixed points in the open disk.

How to obtain such a fundamental domain? A simple way is by doubling a region bounded by any number 
$k$ of non-intersecting geodesics \cite{DB}. Fig.~8a shows this for the case $k=3$. In Fig.~8b we see the
fundamental domain. Half of it is the original (heavily outlined) region, shifted to the right so that
the center of the Poincar\'e disk lies on the geodesic boundary of the region rather than at its
center. The other half is the reflection of this original region
across that geodesic boundary. Thus $2k-2 = 4$ boundaries remain to be identified in pairs, as
indicated in the figure for the top pair. To construct the
isometry that moves one member of such a pair into the other we find the unique 
common normal geodesic H$_2$ (shown for the bottom pair), and its intersection with the limit circle;
these intersections are the fixed 
points of one of the hyperbolic isometries that have this fundamental domain. For example, 
in Fig.~8b the isometry associated with H$_2$ moves
one of the bottom boundaries into the other. Similarly we find $k-2$  other isometries, 
each of them associated with a common normal. After the identification are made these
common normals are smooth closed geodesics that separate an asymptotically AdS region 
from the rest of the manifold. We call such curves horizons. In addition to the 
$k-1$ horizons found this way there is another one, so there is a total of $k$ horizons. The
additional one can be found in the above way from a different fundamental domain, obtained by
reflecting the original region about a different geodesic boundary, but it is
more easily found from the doubling picture, as shown by the H$_3$ in Fig.~8a. 

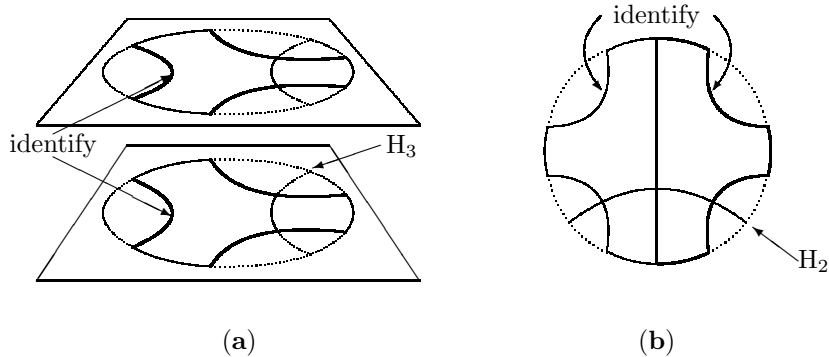
\begin{figure}
\unitlength 0.750mm
\linethickness{0.4pt}
\begin{picture}(140.50,64.00)
\thicklines
\bezier{144}(26.91,35.01)(40.88,29.00)(26.91,22.99)
\bezier{140}(40.69,38.50)(45.65,30.28)(64.67,32.29)
\bezier{140}(40.69,19.54)(45.65,27.72)(64.67,25.63)
\thinlines
\bezier{64}(26.91,35.01)(31.87,37.90)(40.69,38.50)
\bezier{64}(64.67,32.45)(67.52,29.00)(64.67,25.63)
\bezier{64}(26.91,22.99)(31.87,20.10)(40.69,19.54)
\bezier{40}(40.69,38.50)(58.52,39.02)(64.67,32.45)
\bezier{40}(26.91,35.01)(16.44,29.00)(26.91,22.99)
\bezier{40}(40.69,19.54)(58.52,18.98)(64.67,25.63)
\bezier{15}(58.55,36.37)(54.58,34.71)(52.52,32.21)
\bezier{64}(52.52,32.21)(50.47,29.00)(52.68,25.79)
\bezier{15}(52.68,25.79)(54.44,23.23)(58.55,21.63)
\put(26.00,41.00){\line(-2,-3){16.00}}
\thicklines
\put(10.00,17.00){\line(1,0){68.00}}
\thinlines
\put(78.00,17.00){\line(-2,3){16.00}}
\put(62.00,41.00){\line(-1,0){36.00}}
\thicklines
\bezier{144}(26.91,58.71)(40.88,54.00)(26.91,49.29)
\bezier{140}(40.69,61.44)(45.65,55.01)(64.67,56.57)
\bezier{140}(40.69,46.59)(45.65,52.99)(64.67,51.36)
\thinlines
\bezier{64}(26.91,58.71)(31.87,60.97)(40.69,61.44)
\bezier{64}(64.67,56.70)(67.52,54.00)(64.67,51.36)
\bezier{64}(26.91,49.29)(31.87,47.03)(40.69,46.59)
\bezier{40}(40.69,61.44)(58.52,61.85)(64.67,56.70)
\bezier{40}(26.91,58.71)(16.44,54.00)(26.91,49.29)
\bezier{40}(40.69,46.59)(58.52,46.15)(64.67,51.36)
\bezier{15}(58.55,59.78)(54.58,58.47)(52.52,56.51)
\bezier{64}(52.52,56.51)(50.47,54.00)(52.68,51.49)
\bezier{15}(52.68,51.49)(54.44,49.48)(58.55,48.22)
\multiput(26.00,63.40)(-0.12,-0.14){134}{\line(0,-1){0.14}}
\thicklines
\put(10.00,44.60){\line(1,0){68.00}}
\thinlines
\multiput(78.00,44.60)(-0.12,0.14){134}{\line(0,1){0.14}}
\put(62.00,63.40){\line(-1,0){36.00}}
\put(13.00,39.00){\vector(2,-1){21.00}}
\put(13.00,43.00){\vector(2,1){21.00}}
\put(13.00,41.00){\makebox(0,0)[cc]{identify}}
\put(46.00,6.00){\makebox(0,0)[cc]{({\bf a})}}
\thicklines
\put(120.00,20.00){\line(0,1){40.00}}
\thinlines
\put(120.00,6.00){\makebox(0,0)[cc]{({\bf b})}}
\bezier{36}(111.17,57.83)(115.17,60.00)(120.00,60.00)
\thicklines
\bezier{36}(120.00,60.00)(124.83,60.00)(129.00,58.00)
\thinlines
\bezier{20}(129.00,58.00)(137.17,54.17)(139.50,44.17)
\thicklines
\bezier{36}(139.50,44.17)(140.50,40.00)(139.50,35.67)
\bezier{104}(139.50,44.33)(126.83,44.50)(129.00,57.83)
\bezier{36}(120.00,20.00)(124.83,20.00)(129.00,22.00)
\thinlines
\bezier{36}(111.17,22.17)(115.17,20.00)(120.00,20.00)
\bezier{20}(129.00,22.00)(137.17,25.83)(139.50,35.83)
\thicklines
\bezier{104}(139.50,35.67)(126.83,35.50)(129.00,22.17)
\thinlines
\bezier{20}(111.00,58.00)(102.83,54.17)(100.50,44.17)
\bezier{36}(100.50,44.17)(99.50,40.00)(100.50,35.67)
\bezier{104}(100.50,44.33)(113.17,44.50)(111.00,57.83)
\bezier{20}(111.00,22.00)(102.83,25.83)(100.50,35.83)
\bezier{104}(100.50,35.67)(113.17,35.50)(111.00,22.17)
\put(71.00,41.00){\vector(-3,-1){12.00}}
\put(72.00,41.00){\makebox(0,0)[lc]{H$_3$}}
\put(110.00,51.00){\vector(3,-4){0.2}}
\bezier{72}(110.00,64.00)(104.00,58.33)(110.00,51.00)
\put(120.00,64.00){\makebox(0,0)[cc]{identify}}
\put(130.00,51.00){\vector(-3,-4){0.2}}
\bezier{72}(130.00,64.00)(136.00,58.33)(130.00,51.00)
\bezier{156}(104.33,27.33)(120.00,39.33)(135.67,27.33)
\put(145.00,20.50){\vector(-4,3){8.00}}
\put(145.00,20.50){\makebox(0,0)[lc]{H$_2$}}
\end{picture}
\caption{Initial state for an AdS spacetime containing three black holes. 
({\bf a}) Representation by doubling a region on the Poincar\'e disk. The top and 
bottom surfaces are to be glued together along pairs of heavily drawn curves, such as the 
pair labeled ``identify". The resulting topology is that of a pair of pants, with the 
waist and the legs flaring out to infinity at the limit circle. The heavier part of the 
curve H$_3$ becomes a closed geodesic at the narrowest part of a leg.  ({\bf b}) The 
fundamental region, obtained from one of the regions of part ({\bf a}) by 
adding its reflection about the geodesic labeled ``identify" in ({\bf a}). 
In ({\bf b}) only two boundaries remain 
to be identified; the top pair are so labeled. For the bottom pair the minimal 
connecting curve H$_2$ is shown} 
\end{figure}

The topology of the resulting space may be easiest to see in the doubling picture: there are $k$
asymptotically AdS regions, which can be regarded as $k$ punctures (``pants' legs") on a 
2-sphere. With each
asymptotic region there is associated a horizon, namely the geodesic normal to the corresponding
adjacent boundaries of the original region (because it is normal it will become a smooth, 
circular, minimal geodesic after the doubling). On the outside of each horizon the 
geometry is the same as that obtained from the isometry corresponding to that horizon 
alone, so it is exactly the exterior of a BTZ black hole geometry. Therefore the whole 
space contains $k$ black holes, joined together inside each hole's horizon.

\subsubsection{Parameters}

The time-symmetric (zero angular momentum) BTZ black hole in AdS space of a given 
cosmological constant is described by a single parameter, the mass $m$. For an 
initial state of several black holes we have analogously the several masses, and in 
addition the relative positions of the black holes. These are however not all independent.
Consider a $k$ black hole initial state obtained by doubling a simply-connected region 
bounded by $k$ non-intersecting geodesics. Find the $k$ minimal geodesic segments 
$\sigma_i$ between adjacent geodesics.\footnote{Two geodesics of a set 
are adjacent if each has an  end point (at infinity) such that between 
those end points there is no end point of any other geodesic of the set.} 
The parts $s_i$ of the original geodesics between the endpoints of those segments, 
together with the segments $\sigma_i$ themselves, form a 
geodesic $2k$-gon with right-angle corners. Clearly the $\sigma_i$ are half the horizon 
size and hence a measure of the masses, and the $s_i$ may be considered a measure of 
the distances between the black holes. If $2k-3$ of the sides of a $2k$-gon are given,
then the geodesics that will form the $2k-2$ side (orthogonal at the end of the $2k-3$ 
side) and the $2k$ side (orthogonal at the end of the first side) are well-defined.
They have a unique common normal geodesic that forms the $2k-1$ side, hence the whole
polygon is uniquely defined. Thus only $2k-3$ of the $2k$ numbers measuring the masses
and the distances of this type of multi-black-hole are independent. In the case $k=3$
(corresponding to a geodesic hexagon) one can show that alternating sides
(either the three masses or the three distances) can be
{\it arbitrarily} chosen. 
Higher $2k$-gons can be divided by geodesics into hexagons, so at least all 
the masses (or all the distances) can be chosen arbitrarily. (The remaining
$k-3$ parameters may have to satisfy inequalities.)

Composing the $2k$-gon out of geodesic hexagons means, for the doubled surface, that
the multi-black-hole geometry is made out of $k-2$ three-black-hole geometries with
$2k-6$ of the asymptotic AdS regions removed and the horizons glued together
pairwise. In the five-black-hole example of Fig.~9 the three-black-hole parts are
labeled 1, 2, and 3. One asymptotic AdS regions was removed from 1 and 3, and two
such regions are missing from 2. The geometries obtained by doubling this are
however not the most general time-symmetric five-black-hole configuration. For
example, in Fig.~9 the curve separating 
regions 2 and 3 is a closed geodesic. If we cut and re-glue after a hyperbolic isometry 
along this geodesic the geometry is still smooth; the operation amounts to rotating the
top and bottom part of Fig.~9b with respect to each other, as indicated by the arrows.
(In general we can make $k-3$ such re-identifications.) That the result is in
general different after this rotation is shown, for example, by the 
change in angle between the boundary geodesic and another closed geodesic which, 
before the rotation, is indicated by the dotted line in Fig.~9a. 

\begin{figure}
\unitlength 0.89mm
\linethickness{0.4pt}
\begin{picture}(141.55,62.66)(10,0)
\bezier{30}(10.44,35.00)(10.44,48.71)(22.73,56.49)
\bezier{30}(22.73,56.49)(35.00,62.63)(47.27,56.49)
\bezier{30}(47.27,56.49)(59.56,48.71)(59.56,35.00)
\bezier{30}(10.44,35.00)(10.44,21.29)(22.73,13.51)
\bezier{30}(22.73,13.51)(35.00,7.37)(47.27,13.51)
\bezier{30}(47.27,13.51)(59.56,21.29)(59.56,35.00)
\put(10.70,32.50){\oval(3.90,3.40)[r]}
\put(10.70,37.50){\oval(3.90,3.40)[r]}
\bezier{74}(11.17,28.67)(18.33,30.33)(24.33,24.50)
\bezier{70}(24.33,24.50)(29.00,18.67)(27.67,11.50)
\bezier{74}(58.83,28.67)(51.67,30.33)(45.67,24.50)
\bezier{70}(45.67,24.50)(41.00,18.67)(42.33,11.50)
\bezier{74}(58.83,41.33)(51.67,39.67)(45.67,45.50)
\bezier{70}(45.67,45.50)(41.00,51.33)(42.33,58.50)
\bezier{74}(11.17,41.33)(18.33,39.67)(24.33,45.50)
\bezier{70}(24.33,45.50)(29.00,51.33)(27.67,58.50)
\bezier{88}(104.00,60.00)(111.67,56.00)(101.00,48.00)
\bezier{80}(101.00,48.00)(91.00,45.00)(93.00,54.00)
\bezier{32}(89.00,51.00)(92.67,51.00)(90.00,47.00)
\bezier{32}(90.00,47.00)(86.67,43.33)(85.00,46.00)
\bezier{60}(85.00,41.00)(87.00,44.00)(93.00,43.00)
\bezier{80}(93.00,43.00)(99.00,40.33)(97.00,25.00)
\bezier{68}(97.00,25.00)(96.00,15.00)(91.00,14.00)
\bezier{150}(136.00,39.00)(128.67,35.00)(118.00,47.00)
\bezier{120}(118.00,47.00)(112.00,56.33)(118.00,60.00)
\bezier{68}(107.00,61.77)(101.00,60.00)(107.00,58.23)
\bezier{44}(107.00,58.23)(110.67,57.44)(116.00,58.52)
\bezier{52}(116.00,58.52)(120.00,60.00)(116.00,61.48)
\bezier{52}(116.00,61.48)(111.00,62.66)(106.00,61.48)
\thicklines
\bezier{68}(97.00,37.00)(100.00,33.00)(111.00,37.00)
\bezier{72}(111.00,37.00)(121.67,42.00)(118.00,47.00)
\bezier{92}(116.00,25.00)(129.00,32.33)(135.50,25.00)
\bezier{80}(109.00,11.00)(105.67,18.00)(116.00,25.00)
\bezier{68}(134.98,38.30)(131.95,32.00)(134.98,25.70)
\bezier{44}(134.98,25.70)(136.83,22.90)(139.53,26.75)
\bezier{52}(139.53,26.75)(141.55,32.00)(139.53,37.25)
\bezier{52}(139.53,37.25)(137.00,41.45)(134.47,37.25)
\bezier{64}(91.00,14.00)(89.67,9.22)(98.00,7.55)
\bezier{72}(98.00,7.55)(107.00,6.62)(109.00,11.00)
\bezier{24}(95.00,47.30)(96.00,45.00)(93.00,43.00)
\thinlines
\put(92.00,46.00){\makebox(0,0)[cc]{1}}
\put(107.00,44.00){\makebox(0,0)[cc]{2}}
\put(111.00,30.00){\makebox(0,0)[cc]{3}}
\put(24.50,24.50){\line(1,1){21.00}}
\bezier{94}(20.00,42.00)(24.00,35.00)(20.00,28.00)
\put(16.00,35.00){\makebox(0,0)[cc]{1}}
\put(32.00,42.00){\makebox(0,0)[cc]{2}}
\put(41.00,30.00){\makebox(0,0)[cc]{3}}
\bezier{25}(83.17,41.39)(82.43,43.50)(83.17,45.61)
\bezier{25}(83.17,45.61)(84.29,47.02)(85.41,45.61)
\bezier{25}(85.41,45.61)(86.28,43.50)(85.29,41.15)
\bezier{20}(85.29,41.15)(84.17,39.98)(83.17,41.39)
\bezier{30}(88.33,52.00)(88.17,50.33)(90.83,51.83)
\bezier{40}(90.83,51.83)(94.67,54.83)(91.83,54.33)
\bezier{30}(91.83,54.33)(89.83,53.83)(88.33,52.00)
\put(113.33,40.00){\vector(2,1){0.2}}
\bezier{36}(105.00,37.00)(109.67,38.00)(113.33,40.00)
\put(103.33,33.67){\vector(-4,-1){0.2}}
\bezier{44}(113.33,36.67)(107.67,34.00)(103.33,33.67)
\put(35.00,3.00){\makebox(0,0)[cc]{({\bf a})}}
\put(110.00,3.00){\makebox(0,0)[cc]{({\bf b})}}
\thicklines
\bezier{15}(25.00,45.00)(35.00,35.00)(45.00,25.00)
\end{picture}
\caption{A five-black-hole time-symmetric initial state is obtained by doubling the 
region on the Poincar\'e disk in ({\bf a}). Part ({\bf b}) shows a somewhat fanciful picture 
of the result of the doubling, cut off at the flare-outs, which should extend to infinity.} 
\end{figure}
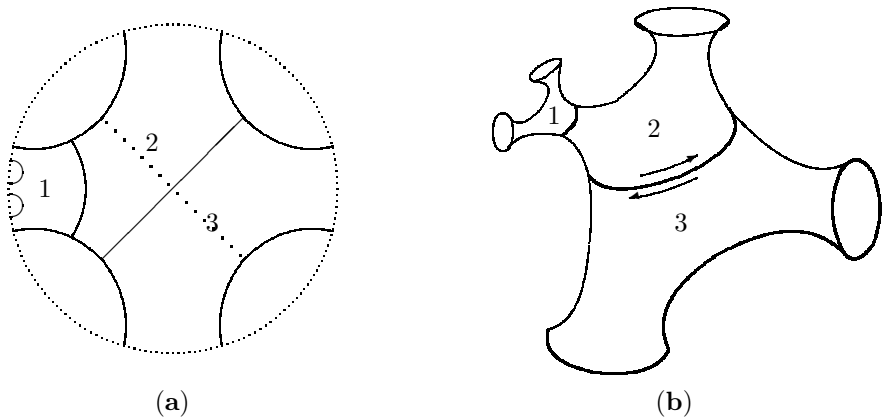

The $2k-3$ distance parameters and the $k-3$ rotation angles describe a $3k-6$-dimensional 
space of $k$-black-hole geometries. Equivalently we may say that a 
$k$-black-hole initial state is given by a fundamental domain bounded by $2k-2$ geodesics
to be identified in pairs by $k-1$ M\"obius transformations. Since each M\"obius 
transformation depends on 3 parameters, and the whole fundamental domain can be moved by
another M\"obius transformation, the number of free parameters is $3k-6$.
Such a space of geometries is known as a Teichm\"uller space, and the
length and twist parameters are known as Fenchel-Nielsen coordinates on this
space \cite{BP}.

Instead of cutting and re-gluing along closed geodesics as in Fig.~9 one can do this
operation on the identification geodesics used in the doubling procedure.
For example, in Fig.~8a on the pair of geodesics marked ``identify" one can
identify each point on the bottom geodesic with one that is moved by a constant
distance along the top geodesic. For the fundamental region this means the
following: so far, whenever two identification geodesics on the boundary of the
fundamental domain were to be identified, it was done by the unique hyperbolic transformation
along the minimal normal geodesic between the identification geodesics. 
If we follow this transformation by a hyperbolic isometry along one of the 
identification geodesics, the two geodesics
will still fit together, and the identified surface will be smooth but with a
difference in global structure (like that produced by the re-gluing in 
Fig.~9). Of course the two transformations
combine into one, and conversely any isometry that maps one identification geodesic
into another can be decomposed into a ``move" along the normal geodesic, and a ``shift"
along a identification geodesic. Since each hyperbolic transformation is a Lorentz
transformation in the embedding picture (Fig.~5) the combination is again
hyperbolic, so no finite fixed points (singularities) occur in this more general 
identification process. 

If we identify with a non-zero shift, there is of course still a minimal
geodesic between the two identified geodesics, but it is no longer
orthogonal to those geodesics. Nevertheless the identified geometry is that
of a black hole. To make the correspondence to the $\phi\rightarrow\phi+2\pi$
identification of Eq (\ref{schwmet}) one would have to change the identification 
geodesics to be normal to the minimal one (which can complicate the fundamental
domain). 

\subsubsection{Fixed points}

It is useful to understand the fixed points at infinity (the limiting circle of the Poincar\'e
disk) of the identifications that glue a black hole geometry out of a 
fundamental domain of AdS space. The fixed points are directly
related to the minimal geodesics associated with the identification, and they
can indicate whether we have a black hole or not: there must be open sets free
of fixed points if the initial data is to be asymptotically AdS. We know that 
the identifications can have some fixed points at infinity, 
but if the fixed points cover all of infinity, there is no place left for an
asymptotically AdS region, and the space is not a black hole space. Thus even in
the relatively simple time-symmetric case it is useful to understand the 
tiling and the fixed points of the M\"obius transformations
associated with the identifications. 

As an example, consider again
the three-black-hole case. Let $a$ and $b$ be the identifications of the top and 
the bottom pair of geodesics of a figure like 8b. Then the free group generated 
by these, that is, any ``word" formed from $a,\,b$ and their inverses $A,\,B$ is also 
an identification. Since the identified geometry is everywhere smooth, none of these can 
have a fixed point in the finite part of the disk, so all fixed points 
must lie on the 
limit circle. The pattern of fixed points is characteristic of the identifications
and constitutes a kind of hologram \cite{Suss} of the multi-black-hole spacetime.

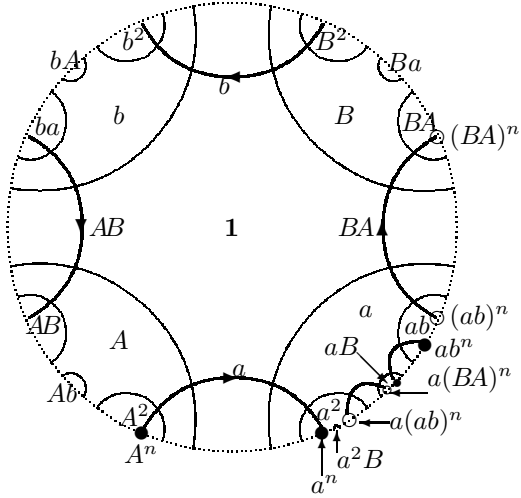
\begin{figure} 
\unitlength 1.00mm
\linethickness{0.4pt}
\begin{picture}(69.85,68.58)(20,15)
\bezier{49}(20.15,50.00)(20.15,66.66)(35.08,76.12)
\bezier{48}(35.08,76.12)(50.00,83.58)(64.92,76.12)
\bezier{49}(64.92,76.12)(79.85,66.66)(79.85,50.00)
\bezier{49}(20.15,50.00)(20.15,33.34)(35.08,23.88)
\bezier{48}(35.08,23.88)(50.00,16.42)(64.92,23.88)
\bezier{49}(64.92,23.88)(79.85,33.34)(79.85,50.00)
\bezier{88}(55.19,20.56)(53.33,30.93)(61.11,39.07)
\bezier{84}(61.11,39.07)(68.70,46.48)(79.44,44.81)
\bezier{32}(59.25,21.67)(58.17,24.83)(61.50,27.33)
\bezier{32}(61.50,27.33)(65.58,29.00)(68.00,26.00)
\bezier{12}(69.75,27.42)(69.00,28.83)(70.00,29.92)
\bezier{32}(78.33,40.75)(75.17,41.83)(72.67,38.50)
\bezier{32}(72.67,38.50)(71.00,34.42)(74.00,32.00)
\bezier{12}(72.58,30.25)(71.17,31.00)(70.08,30.00)
\put(65.67,24.33){\circle{1.67}}
\put(70.67,28.33){\circle{1.00}}
\put(62.00,22.67){\circle*{1.67}}
\bezier{8}(63.50,23.00)(63.50,24.17)(64.33,23.33)
\put(67.83,38.83){\makebox(0,0)[cc]{$a$}}
\put(63.00,25.80){\makebox(0,0)[cc]{$a^2$}}
\put(74.67,37.00){\makebox(0,0)[cc]{$ab$}}
\bezier{88}(79.44,55.19)(69.07,53.33)(60.93,61.11)
\bezier{84}(60.93,61.11)(53.52,68.70)(55.19,79.44)
\bezier{32}(78.33,59.25)(75.17,58.17)(72.67,61.50)
\bezier{32}(72.67,61.50)(71.00,65.58)(74.00,68.00)
\bezier{12}(72.58,69.75)(71.17,69.00)(70.08,70.00)
\bezier{32}(59.25,78.33)(58.17,75.17)(61.50,72.67)
\bezier{32}(61.50,72.67)(65.58,71.00)(68.00,74.00)
\bezier{12}(69.75,72.58)(69.00,71.17)(70.00,70.08)
\bezier{88}(44.81,79.44)(46.67,69.07)(38.89,60.93)
\bezier{84}(38.89,60.93)(31.30,53.52)(20.56,55.19)
\bezier{32}(40.75,78.33)(41.83,75.17)(38.50,72.67)
\bezier{32}(38.50,72.67)(34.42,71.00)(32.00,74.00)
\bezier{12}(30.25,72.58)(31.00,71.17)(30.00,70.08)
\bezier{32}(21.67,59.25)(24.83,58.17)(27.33,61.50)
\bezier{32}(27.33,61.50)(29.00,65.58)(26.00,68.00)
\bezier{12}(27.42,69.75)(28.83,69.00)(29.92,70.00)
\bezier{88}(20.56,44.81)(30.93,46.67)(39.07,38.89)
\bezier{84}(39.07,38.89)(46.48,31.30)(44.81,20.56)
\bezier{32}(21.67,40.75)(24.83,41.83)(27.33,38.50)
\bezier{32}(27.33,38.50)(29.00,34.42)(26.00,32.00)
\bezier{12}(27.42,30.25)(28.83,31.00)(29.92,30.00)
\bezier{32}(40.75,21.67)(41.83,24.83)(38.50,27.33)
\bezier{32}(38.50,27.33)(34.42,29.00)(32.00,26.00)
\bezier{12}(30.25,27.42)(31.00,28.83)(30.00,29.92)
\put(65.00,65.00){\makebox(0,0)[cc]{$B$}}
\put(35.00,65.00){\makebox(0,0)[cc]{$b$}}
\put(35.00,35.00){\makebox(0,0)[cc]{$A$}}
\put(75.00,64.00){\makebox(0,0)[cc]{$B\!A$}}
\put(73.00,72.00){\makebox(0,0)[cc]{$Ba$}}
\put(63.30,75.00){\makebox(0,0)[cc]{$B^2$}}
\put(37.00,75.00){\makebox(0,0)[cc]{$b^2$}}
\put(30.00,71.00){\makebox(0,0)[rb]{$bA$}}
\put(25.50,63.50){\makebox(0,0)[cc]{$ba$}}
\put(25.00,37.00){\makebox(0,0)[cc]{$A\!B$}}
\put(29.50,28.00){\makebox(0,0)[rc]{$Ab$}}
\put(37.00,25.00){\makebox(0,0)[cc]{$A^2$}}
\put(38.00,22.67){\circle*{1.67}}
\put(72.00,29.33){\circle*{1.00}}
\put(75.67,34.33){\circle*{1.67}}
\put(77.33,38.00){\circle{1.67}}
\put(77.33,62.00){\circle{1.67}}
\thicklines
\bezier{64}(38.00,23.00)(42.04,30.00)(50.00,30.00)
\bezier{64}(50.00,30.00)(58.15,30.00)(62.00,23.00)
\bezier{44}(65.17,25.17)(65.33,30.67)(70.17,28.67)
\bezier{44}(74.83,34.83)(69.33,34.67)(71.33,29.83)
\bezier{64}(77.00,62.00)(70.00,57.96)(70.00,50.00)
\bezier{64}(70.00,50.00)(70.00,41.85)(77.00,38.00)
\bezier{64}(23.00,38.00)(30.00,42.04)(30.00,50.00)
\bezier{64}(30.00,50.00)(30.00,58.15)(23.00,62.00)
\bezier{64}(62.00,77.00)(57.96,70.00)(50.00,70.00)
\bezier{64}(50.00,70.00)(41.85,70.00)(38.00,77.00)
\put(50.00,30.00){\vector(1,0){1.00}}
\put(70.00,50.00){\vector(0,1){1.00}}
\put(50.00,70.00){\vector(-1,0){1.00}}
\put(30.00,50.00){\vector(0,-1){1.00}}
\thinlines
\put(51.00,31.00){\makebox(0,0)[cc]{$a$}}
\put(49.00,69.00){\makebox(0,0)[cc]{$b$}}
\put(69.00,50.00){\makebox(0,0)[rc]{$B\!A$}}
\put(31.00,50.00){\makebox(0,0)[lc]{$A\!B$}}
\put(66.70,33.30){\vector(1,-1){4.00}}
\put(67.00,33.00){\makebox(0,0)[rb]{$aB$}}
\put(62.00,17.00){\vector(0,1){5.00}}
\put(62.40,17.00){\makebox(0,0)[ct]{$a^n$}}
\put(64.00,20.00){\vector(0,1){3.00}}
\put(64.00,20.80){\makebox(0,0)[lt]{$a^2B$}}
\put(71.00,24.00){\vector(-1,0){4.00}}
\put(71.00,24.00){\makebox(0,0)[lc]{$a(ab)^n$}}
\put(76.00,28.00){\vector(-1,0){5.00}}
\put(76.00,28.00){\makebox(0,0)[lb]{$a(B\!A)^n$}}
\put(79.00,38.00){\makebox(0,0)[lc]{$(ab)^n$}}
\put(79.00,62.00){\makebox(0,0)[lc]{$(B\!A)^n$}}
\put(38.00,21.00){\makebox(0,0)[ct]{$A^n$}}
\put(77.00,34.00){\makebox(0,0)[lc]{$ab^n$}}
\put(50.00,50.00){\makebox(0,0)[cc]{\bf 1}}
\end{picture}
\caption{Part of the tiling of the Poincar\'e disk obtained by ``unwrapping" a
three-black-hole initial geometry as in Fig.~8. A fundamental domain
{\bf 1} is imaged by combinations of identification maps $a$ and $b$ and their
inverses $A=a^{-1},\,B=b^{-1}$. Repeating $n$ times a map such as $ab$ leads to
a point $(ab)^n$ on the limit circle, in the limit $n\rightarrow\infty$. 
Some geodesics (``horizons") connecting such a limit point and its inverse 
limit (such as $a(ab)^n$ and $a(ab)^{-n}=a(BA)^n$) are shown as heavy curves} 
\label{fig10} 
\end{figure} 

In Fig.~10 the initial fundamental domain
is denoted by {\bf 1}. The identifications are given by hyperbolic M\"obius 
transformations $a,\,b$, with inverses $A,\,B$ that connect the top and 
the bottom boundaries, respectively. Any ``word" made up of these four
letters is, first, also an identification. Secondly each word can be used to label a tile, 
because each tile is some image, $a${\bf 1}, $A${\bf 1}, $aB${\bf 1},\dots of the initial 
domain {\bf 1}, shown simply as $a,\,A,\,aB,$ \dots in the Figure. Finally, there is a 
closed minimal geodesic associated with each pair of identified boundaries, hence
each word also corresponds to a geodesic.\footnote{In this connection we regard a
word, its inverse, and the permuted word as equal, in order to have a unique
correspondence to geodesics; see \cite{sor}.} (For example, $Ba$ connects $(Ab)^n$
to $(Ba)^n$.) Horizons are special geodesics that bound asymptotically 
AdS regions. Some of these are shown by the heavy curves. The 
ones that cut through the basic domain are labeled by the isometries that
leave them invariant, $a,\,b$, and $AB=BA$. The words for the
other horizons are obtained from these by conjugation, for example the
horizon connecting the points labeled $a(ab)^n$ and $a(BA)^n$ is ``called"
by the word $a(BA)A$.

Every words is a hyperbolic isometry, hence has two fixed points on the
Poincar\'e limit circle. We can find the fixed points by applying the word
(or its inverse) many times to any finite region, because in the limit 
the images will converge to a point on the limit circle (see, 
for example, the equidistant curves in Fig.~2). Some of these fixed
points are shown by open and by filled circles in the Figure, and labeled
by an $n$th power, where the limit $n\rightarrow\infty$ is understood.
The two fixed points of a hyperbolic transformation define a geodesic that
ends at them, and that is the minimal geodesic along which the transformation
acts.

Because the infinity side of a horizon is isometric to the asymptotic region
of a single black hole, there are no fixed points on that side of the
horizon. (Cf.~Fig.~2, where the only fixed points of the horizon isometry
$\phi\rightarrow\phi+$ const are on the horizon $r=\ell$.) Between two
different horizons (between open and filled circles of the Figure) 
there will however be further horizons, with fixed points at their ends.
Thus the set of fixed points for multi-black-holes has the fractal structure
of a Cantor set.

By contrast, for some identifications the fixed points are everywhere dense
on the limit circle. This happens, for example, if we try to build, by
analogy to the multi-black-hole construction, a geometry containing three
$m=0$ black holes. The tiles, analogs of those of Fig.~10, would be ``ideal"
quadrilaterals, that is, each tile is a geodesic polygon whose four corners 
lie on the limit circle. This space is smooth and contains three ends of the
type shown in the second column of Fig.~7 (instead of the ``legs" in such
pictures as Fig.~9); but since there is then no fixed-point-free region on
the limit circle, this space is not asymptotically AdS and hence does not
contain BTZ-type black holes.

\subsection{Other Topologies}

It is well known that time-symmetric AdS initial states, that is,
spaces of constant negative curvature, admit a large variety of topologies.
In the context of (orientable) black hole spaces one can construct all of these
out of pieces of the three-black-hole space as in Fig.~7. These pieces are:
three BTZ-exteriors, that is, the regions outside each of the three horizons;
and one region interior to the horizons. The interior piece is sometimes
called the ``convex core" or ``trousers."\footnote{Previously we have used
the image of flared pants' legs for the asymptotically AdS regions, which
need to be cut off to obtain the core, so it would be more consistent to
call the latter ``cut-offs" or ``shorts," but we will use ``trousers."} Fig.~11
shows how other topologies can be constructed out of these pieces. 

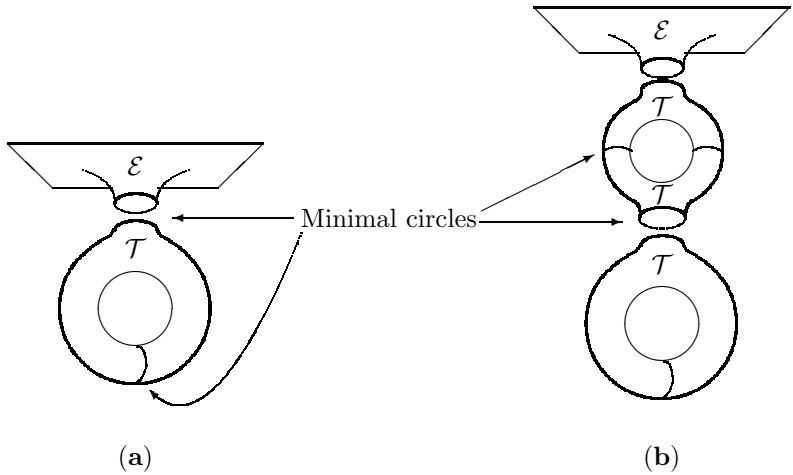
\begin{figure}
\unitlength 1.00mm
\linethickness{0.4pt}
\begin{picture}(117.00,70.00)(10,0)
\put(30.00,30.00){\circle{10.00}}
\put(27.00,46.00){\line(-1,0){8.00}}
\put(19.00,46.00){\line(-1,1){6.00}}
\thicklines
\put(13.00,52.00){\line(1,0){17.00}}
\thinlines
\put(33.00,46.00){\line(1,0){8.00}}
\put(41.00,46.00){\line(1,1){6.00}}
\thicklines
\put(47.00,52.00){\line(-1,0){17.00}}
\put(30.00,49.00){\makebox(0,0)[cc]{$\cal E$}}
\bezier{32}(19.78,30.00)(19.78,34.14)(22.71,37.07)
\bezier{32}(39.78,30.00)(39.78,34.14)(36.85,37.07)
\bezier{32}(19.78,30.00)(19.78,25.86)(22.71,22.93)
\bezier{32}(22.71,22.93)(25.64,20.00)(29.78,20.00)
\bezier{32}(39.78,30.00)(39.78,25.86)(36.85,22.93)
\bezier{32}(36.85,22.93)(33.92,20.00)(29.78,20.00)
\bezier{8}(27.13,44.00)(27.13,44.53)(27.95,44.91)
\bezier{8}(27.95,44.91)(28.77,45.29)(29.94,45.29)
\bezier{8}(32.75,44.00)(32.75,44.53)(31.93,44.91)
\bezier{8}(31.93,44.91)(31.10,45.29)(29.94,45.29)
\thinlines
\bezier{8}(27.13,44.00)(27.13,43.47)(27.95,43.09)
\bezier{8}(27.95,43.09)(28.77,42.71)(29.94,42.71)
\bezier{8}(32.75,44.00)(32.75,43.47)(31.93,43.09)
\bezier{8}(31.93,43.09)(31.10,42.71)(29.94,42.71)
\bezier{24}(32.75,44.00)(32.75,47.00)(37.00,48.40)
\bezier{24}(27.21,44.00)(27.21,47.00)(22.96,48.40)
\thicklines
\bezier{8}(27.13,40.40)(27.13,40.93)(27.95,41.31)
\bezier{8}(27.95,41.31)(28.77,41.69)(29.94,41.69)
\bezier{8}(32.75,40.40)(32.75,40.93)(31.93,41.31)
\bezier{8}(31.93,41.31)(31.10,41.69)(29.94,41.69)
\bezier{16}(36.84,37.07)(35.64,38.40)(33.38,39.60)
\bezier{4}(33.38,39.60)(32.98,39.87)(32.71,40.53)
\bezier{16}(22.84,37.07)(24.04,38.40)(26.31,39.60)
\bezier{4}(26.31,39.60)(26.71,39.87)(26.98,40.53)
\thinlines
\bezier{16}(30.18,24.93)(31.24,25.07)(31.38,22.40)
\bezier{12}(31.38,22.40)(30.98,20.53)(30.04,20.13)
\put(97.00,64.00){\line(-1,0){8.00}}
\put(89.00,64.00){\line(-1,1){6.00}}
\thicklines
\put(83.00,70.00){\line(1,0){17.00}}
\put(117.00,70.00){\line(-1,0){17.00}}
\thinlines
\put(103.00,64.00){\line(1,0){8.00}}
\put(111.00,64.00){\line(1,1){6.00}}
\put(100.00,67.00){\makebox(0,0)[cc]{$\cal E$}}
\thicklines
\bezier{8}(97.13,62.00)(97.13,62.53)(97.95,62.91)
\bezier{8}(97.95,62.91)(98.77,63.29)(99.94,63.29)
\bezier{8}(102.75,62.00)(102.75,62.53)(101.93,62.91)
\bezier{8}(101.93,62.91)(101.10,63.29)(99.94,63.29)
\thinlines
\bezier{8}(97.13,62.00)(97.13,61.47)(97.95,61.09)
\bezier{8}(97.95,61.09)(98.77,60.71)(99.94,60.71)
\bezier{8}(102.75,62.00)(102.75,61.47)(101.93,61.09)
\bezier{8}(101.93,61.09)(101.10,60.71)(99.94,60.71)
\bezier{24}(102.75,62.00)(102.75,65.00)(107.00,66.40)
\bezier{24}(97.21,62.00)(97.21,65.00)(92.96,66.40)
\put(100.00,51.00){\circle{8.00}}
\thicklines
\bezier{8}(97.13,59.00)(97.13,59.53)(97.95,59.91)
\bezier{8}(97.95,59.91)(98.77,60.29)(99.94,60.29)
\bezier{8}(102.75,59.00)(102.75,59.53)(101.93,59.91)
\bezier{8}(101.93,59.91)(101.10,60.29)(99.94,60.29)
\bezier{8}(96.84,42.05)(96.84,42.64)(97.74,43.05)
\bezier{8}(97.74,43.05)(98.64,43.47)(99.93,43.47)
\bezier{8}(103.03,42.05)(103.03,42.64)(102.13,43.05)
\bezier{8}(102.13,43.05)(101.21,43.47)(99.93,43.47)
\thinlines
\bezier{8}(96.84,42.05)(96.84,41.47)(97.74,41.05)
\bezier{8}(97.74,41.05)(98.64,40.63)(99.93,40.63)
\bezier{8}(103.03,42.05)(103.03,41.47)(102.13,41.05)
\bezier{8}(102.13,41.05)(101.21,40.63)(99.93,40.63)
\thicklines
\bezier{12}(102.81,58.97)(102.81,57.90)(104.09,57.44)
\bezier{36}(104.09,57.44)(107.94,55.38)(107.94,51.03)
\bezier{36}(104.09,44.40)(107.94,46.67)(107.94,51.03)
\bezier{20}(104.09,44.40)(103.00,44.00)(103.00,42.26)
\bezier{12}(97.19,58.97)(97.19,57.90)(95.91,57.44)
\bezier{36}(95.91,57.44)(92.06,55.38)(92.06,51.03)
\bezier{36}(95.91,44.40)(92.06,46.67)(92.06,51.03)
\bezier{20}(95.91,44.40)(97.00,44.00)(97.00,42.26)
\thinlines
\put(100.00,28.00){\circle{10.00}}
\thicklines
\bezier{32}(89.78,28.00)(89.78,32.14)(92.71,35.07)
\bezier{32}(109.78,28.00)(109.78,32.14)(106.85,35.07)
\bezier{32}(89.78,28.00)(89.78,23.86)(92.71,20.93)
\bezier{32}(92.71,20.93)(95.64,18.00)(99.78,18.00)
\bezier{32}(109.78,28.00)(109.78,23.86)(106.85,20.93)
\bezier{32}(106.85,20.93)(103.92,18.00)(99.78,18.00)
\bezier{8}(97.13,38.40)(97.13,38.93)(97.95,39.31)
\bezier{8}(97.95,39.31)(98.77,39.69)(99.94,39.69)
\bezier{8}(102.75,38.40)(102.75,38.93)(101.93,39.31)
\bezier{8}(101.93,39.31)(101.10,39.69)(99.94,39.69)
\bezier{16}(106.84,35.07)(105.64,36.40)(103.38,37.60)
\bezier{4}(103.38,37.60)(102.98,37.87)(102.71,38.53)
\bezier{16}(92.84,35.07)(94.04,36.40)(96.31,37.60)
\bezier{4}(96.31,37.60)(96.71,37.87)(96.98,38.53)
\thinlines
\bezier{16}(100.18,22.93)(101.24,23.07)(101.38,20.40)
\bezier{12}(101.38,20.40)(100.98,18.53)(100.04,18.13)
\bezier{20}(92.14,50.97)(94.02,52.0)(95.91,50.97)
\bezier{20}(107.79,50.97)(105.91,52.0)(104.02,50.97)
\put(100.00,57.00){\makebox(0,0)[cc]{$\cal T$}}
\put(100.00,44.00){\makebox(0,0)[cb]{$\cal T$}}
\put(100.00,36.00){\makebox(0,0)[cc]{$\cal T$}}
\put(30.00,38.00){\makebox(0,0)[cc]{$\cal T$}}
\put(51.00,42.00){\vector(-1,0){16.00}}
\put(52.00,42.00){\makebox(0,0)[lc]{Minimal circles}}
\put(76.00,43.00){\vector(2,1){15.00}}
\put(75.74,41.49){\vector(1,0){19.00}}
\put(32.00,19.00){\vector(-1,1){0.2}}
\bezier{172}(52.00,40.00)(40.00,10.51)(32.00,19.00)
\put(30.00,10.00){\makebox(0,0)[cc]{({\bf a})}}
\put(100.00,10.00){\makebox(0,0)[cc]{({\bf b})}}
\end{picture}
\caption{Examples of the class of spaces considered here, constructed by sewing
together one or several trousers and one asymptotic region. The latter looks
asymptotically flat in this topological picture, but metrically it has constant
negative curvature everywhere, just like the trousers}
\end{figure}

The resulting
geometry is smooth if we choose the freely specifiable mass parameters of
each exterior or core to match those of its neighbors at the connection 
horizons\footnote{The horizons along which the legs were cut off from
the cores may no longer be horizons of the space-time if the cores are
re-assembled differently. Nonetheless, in the present section we will still 
call them by that name.}: the intrinsic geometries then match, and
the extrinsic geometries match because the horizons are geodesics. Conversely
we can decompose a given $k$-black-hole initial geometry of genus $g$ into 
asymptotic regions and
trousers by cutting it along minimal circles of different and non-trivial
homotopy types. We can choose $3g+2k-3$ such circles that divide the
space into $k$ exteriors $\cal E$ and $2g+k-2$ trousers $\cal T$. Fig.~11
illustrates the construction for $k=1$ and $g=1$ (left) resp.\ $g=2$ (right).

A fundamental region on the Poincar\'e disk, and hence the M\"obius transformations
that implement the identifications, can be constructed for these spaces in
a similar way, by putting together geodesic, right-angle octagons 
representing trousers and analogous asymptotic regions. For example, the $k=1,\,
g=1$ geometry can be represented by identifying two of the horizon geodesics
of a trousers octagon and adding an exterior to the third. The resulting
fundamental domain is bounded by geodesics, but it is not 
unique. We can cut it into pieces and re-assemble it in a different way 
\cite{DBS}, or we can cut the original space along some geodesics (not
necessarily those of the trousers decomposition) only until it becomes 
one simply connected piece. If we can lay these geodesics so that they
start and end at infinity and therefore do not cross we obtain a simple
fundamental region bounded only by complete geodesics. Figure 12a shows 
the two geodesic cuts necessary for the case of our example of Fig.~11a, and 
the fundamental domain so obtained is seen in Fig.~12b.
The pattern of tiling for this case is identical to that of Fig.~10, but the 
labeling is different. For example, rather than three horizon words there
is only one, $abAB$, corresponding to the existence of only one horizon
in the identified manifold.

\begin{figure}
\unitlength 0.80mm
\linethickness{0.4pt}
\begin{picture}(121.27,66.23)
\bezier{44}(76.73,33.00)(76.73,45.43)(87.87,52.49)
\bezier{40}(87.87,52.49)(99.00,58.06)(110.13,52.49)
\bezier{44}(110.13,52.49)(121.27,45.43)(121.27,33.00)
\bezier{44}(76.73,33.00)(76.73,20.57)(87.87,13.51)
\bezier{40}(87.87,13.51)(99.00,7.94)(110.13,13.51)
\bezier{44}(110.13,13.51)(121.27,20.57)(121.27,33.00)
\thicklines 
\bezier{88}(117.28,45.81)(110.69,40.84)(110.69,33.00)
\bezier{88}(117.28,20.19)(110.69,25.16)(110.69,33.00)
\bezier{88}(80.72,45.81)(87.31,40.84)(87.31,33.00)
\bezier{88}(80.72,20.19)(87.31,25.16)(87.31,33.00)
\bezier{64}(90.66,53.65)(93.28,47.92)(99.00,47.92)  
\bezier{64}(107.34,53.65)(104.72,47.92)(99.00,47.92) 
\bezier{64}(90.66,12.35)(93.28,18.08)(99.00,18.08)  
\bezier{64}(107.34,12.35)(104.72,18.08)(99.00,18.08)
\thinlines
\put(99.00,18.08){\vector(0,1){29.85}}
\put(87.43,33.00){\vector(1,0){23.13}}
\put(40.00,28.67){\circle{12.00}}
\thicklines
\bezier{64}(21.67,30.00)(21.67,40.23)(30.84,46.04)
\bezier{64}(49.16,46.04)(58.33,40.23)(58.33,30.00)  
\bezier{64}(21.67,30.00)(21.67,19.77)(30.84,13.96)   
\bezier{60}(30.84,13.96)(40.00,9.38)(49.16,13.96)   
\bezier{64}(49.16,13.96)(58.33,19.77)(58.33,30.00)  
\bezier{48}(30.84,46.04)(36.67,50.00)(37.00,54.00)
\bezier{68}(37.00,54.00)(37.67,63.00)(28.00,63.00)
\bezier{48}(49.16,46.04)(43.33,50.00)(43.00,54.00)
\bezier{68}(43.00,54.00)(42.33,63.00)(52.00,63.00)
\thinlines
\bezier{64}(27.59,63.00)(27.59,61.40)(33.80,60.49)
\bezier{60}(33.80,60.49)(40.00,59.77)(46.20,60.49)  
\bezier{64}(46.20,60.49)(52.41,61.40)(52.41,63.00)
\bezier{64}(27.59,63.00)(27.59,64.60)(33.80,65.51)  
\bezier{60}(33.80,65.51)(40.00,66.23)(46.20,65.51)  
\bezier{64}(46.20,65.51)(52.41,64.60)(52.41,63.00)
\thicklines
\bezier{68}(40.00,35.00)(45.00,34.00)(42.00,45.00)
\bezier{64}(42.00,45.00)(40.00,55.33)(42.00,60.00)
\bezier{14}(40.00,35.00)(34.67,35.00)(37.50,49.17)
\bezier{10}(37.50,49.17)(38.83,56.33)(38.17,60.17)
\bezier{32}(38.17,60.17)(36.83,64.83)(33.67,65.50)  
\thinlines
\bezier{14}(40.24,22.39)(38.86,22.39)(38.07,19.69)  
\bezier{10}(38.07,19.69)(37.45,16.98)(38.07,14.28)  
\bezier{14}(38.07,14.28)(38.86,11.56)(40.24,11.56)
\bezier{34}(40.24,22.39)(41.63,22.39)(42.42,19.69)
\bezier{30}(42.42,19.69)(43.03,16.98)(42.42,14.28)
\bezier{34}(42.42,14.28)(41.63,11.56)(40.24,11.56)
\put(95.67,33.33){\makebox(0,0)[cb]{$a$}}
\put(99.33,39.67){\makebox(0,0)[lc]{$b$}}
\put(42.87,17.00){\makebox(0,0)[lc]{$a$}}
\put(46.33,28.33){\makebox(0,0)[lc]{$b$}}
\put(44.33,33.00){\vector(-1,1){0.17}}
\put(41.76,21.44){\vector(-1,3){0.2}}
\put(40.00,3.00){\makebox(0,0)[cc]{({\bf a})}}
\put(100.00,3.00){\makebox(0,0)[cc]{({\bf b})}}
\end{picture}
\caption{Construction of a BTZ exterior with toroidal interior. Rather than
cutting the geometry shown in part ({\bf a}) by minimal geodesics as in
Fig.~11, the cuts, shown by the heavy lines, are chosen to reach infinity
and divide this into four regions. Thus one obtains the fundamental domain
shown in part ({\bf b}). The identifications $a$ and $b$ are shown as
arrows. The lines of these arrows are also the minimal geodesics between 
the lines that are to be identified. In the identified manifold these are
closed geodesics, as shown in part ({\bf a}). The possible geometries are
characterized by the lengths of these two closed geodesics, and the angle
between them (shown here as $90^\circ$)}
\end{figure}
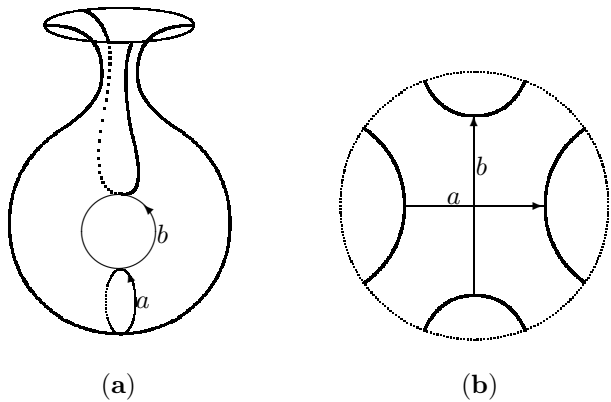

Out of an even number of trousers only one can construct locally AdS initial
data that contain
no asymptotically AdS region at all. Such compact spaces can be interpreted
as closed universes, and for lack of other physical content they can be
considered to contain several black holes, associated with the horizons
that were glued together in the construction. (Of course these horizons
and black holes are only analogies, for example there are no observers that
see them as black, i.e.\ of infinite redshift.) 

Our reasoning about the number of parameters that specify a $k$-black-hole
geometry can be generalized to the case that the internal geometry has
genus $g$. If we cut off two asymptotic AdS regions and identify the two
horizons that go with them, we decrease $k$ by two and increase $g$ by one.
The number of parameters does not change: we lose one mass parameter,
since the masses of the two horizons that will be identified have to be
equal, but we gain one rotation parameter which specifies with what shift
the horizons are to be identified. Thus from the formula in section 2.4
we find that the (orientable) time-symmetric initial states, of genus $g$ 
and $k$ asymptotic AdS regions, form a $(6g+3k-6)$-dimensional 
Teichm\"uller space. 
If this number is non-positive, no state of that type is possible.
(However, the formula cannot be applied to the time-symmetric BTZ 
initial state itself:
it has one free parameter, the mass $m$, but no integral value of $k$
makes the formula valid; the BTZ state is not a multi-black-hole geometry
in the sense of this section.)
For example, if we want a single exterior region ($k=1$) we need a genus
of at least $g=1$ (Fig.~12). Here the number of parameters is $6g+3k-6=3$,
for example the minimal distances (lengths of closed geodesics $a$ and $b$) 
for each of the two identifications, and the angle between these
geodesics. It is clear from the figure that these distances
must be large enough, and the angle close enough to a right angle, 
that an asymptotic region remains in Fig.~12b. (If the 
geodesics crossed and formed
a quadrilateral, there would be an angle deficit at the crossing point,
which could be interpreted as a toroidal universe that is not empty,
but contains one point particle.)

The formula for the number of free parameters tells us that there is 
no time-symmetric torus ($k=0,\,g=1$) initial state. However, 
all topologies of higher genus or with at least one asymptotic AdS region 
do occur; and the spatial torus topology does
occur among all locally AdS spacetimes, for example as Eq~(\ref{schwmet})
for $r^2<\ell^2$ with $\phi$ and $t$ periodically identified --- the analog 
of a closed Kantowski-Sachs universe. 

\section{Time Development}

The identifications used on a time-symmetric surface of AdS space to 
generate black hole and other initial values have a unique extension 
to all of AdS space, and thus define a unique time development (even 
beyond any Cauchy horizon). A fundamental domain in 
3-dimensional AdS space can be generated by extending normal 
timelike geodesics from the geodesic boundaries of the two-dimensional 
fundamental domain on the initial surface. Due to the negative curvature 
of AdS space such {\it timelike} geodesics accelerate towards each other 
and will eventually cross. Such crossing of fundamental domain boundaries 
is the space-time analog of a conical singularity. A prototype of this is the
``non-Hausdorff singularity" of Misner space \cite{Mis}. Although not a curvature singularity, 
these points are considered not to be part of the space-time. This in turn 
provides an end of \scri\ and hence the possibility of a black hole horizon.

A metric for the time development of the finite part of any multi-black-hole
or multiply-connected time-symmetric initial geometry is provided by Eq (\ref{cmc})
when we replace the expression in the bracket by the initial multi-black-hole
metric. The result is a metric adapted to free-fall observers, and it shows
that they all reach the singularity after the same proper time, $\tau=\pi\ell/2$,
when the $\cos^2$ factor vanishes. (This can be seen geometrically from Fig.~4a, where 
geodesics are intersections with planes through the origin, and the 
collapse time is one quarter of the period around the hyperboloid.)
But these coordinates do not cover the time development of conformal
infinity (cf.\ the dotted curves in Fig.~2).

A more complete picture emerges from the continuation of the 
identification group to AdS spacetime, for example
via the embedding of AdS space according to Eq (\ref{ads}). In 
the embedding of the initial surface in the 3-dimensional Minkowski space 
$V=0$, each identification corresponds to a Lorentz ``rotation" about 
some (spacelike) axis $A$. This is uniquely extended to an SO(2,2) 
``rotation" of the four-dimensional embedding space by requiring that 
the $V$-axis also remain invariant; that is, we rotate by the same 
hyperbolic angle about the $A,V$ plane. This plane intersects the AdS 
space (\ref{ads}) in a spacelike geodesic of fixed points. All such 
geodesics from all the identifications are to be considered 
singularities after the identifications are made, so they are not points 
in the identified spacetime.

Three-dimensional pictures that include conformal infinity and all of the
singularities can be had in sausage and in stereographic coordinates,
Eqs (\ref{sausge}) and (\ref{stm}). Because all timelike geodesics starting 
normally on a time-symmetric initial surface collapse together to a point C,
all the totally geodesic boundaries of the fundamental 
domain also meet at C, forming a tent-like structure with a tip 
at C. Their intersections may be timelike or spacelike. If an 
intersection is timelike, the sides typically intersect there at a right 
angle ``corner," and the intersection passes through the initial surface. If 
the initial geometry is smooth, such intersections are 
innocuous.\footnote{We have not encountered such corners in our pictures, 
but they must appear in spaces composed only of trousers, for example in 
the time development of a $k=0,\,g=2$ surface that can be represented by 
a right-angled octagon on the Poincar\'e disk, as in Fig.~3b of \cite{DBS}.} 
Spacelike intersections are called ``folds" of 
the tent, and they are the geodesics of fixed points, which likewise meet 
at C.\footnote{The reason that corners can be regular and but folds are 
not is that four corners can be put together to make a line without angle 
deficit, but no finite number of folds can eliminate the 
Misner-space singularity.}

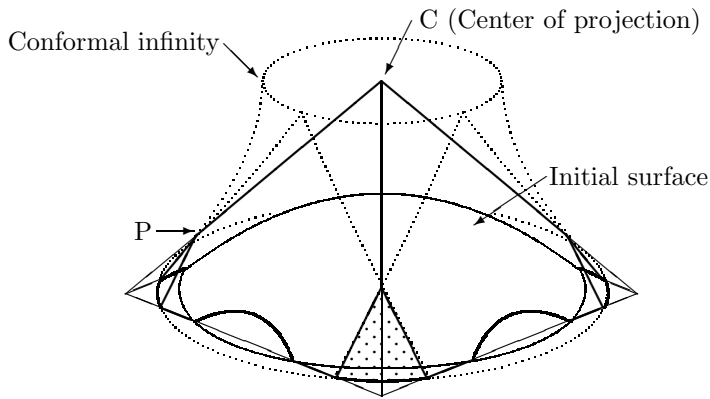
\begin{figure}
\unitlength 1.00mm
\linethickness{0.4pt}
\begin{picture}(84.04,55.00)(-7,15)
\bezier{30}(20.15,32.00)(20.15,38.59)(35.08,42.34)
\bezier{30}(64.92,42.34)(79.85,38.59)(79.85,32.00)
\bezier{30}(20.15,32.00)(20.15,25.41)(35.08,21.66)
\bezier{40}(35.08,21.66)(50.00,18.71)(64.92,21.66)
\bezier{30}(64.92,21.66)(79.85,25.41)(79.85,32.00)
\bezier{14}(34.17,60.00)(34.17,63.17)(42.09,64.97)
\bezier{20}(42.09,64.97)(50.00,66.39)(57.91,64.97)
\bezier{14}(57.91,64.97)(65.83,63.17)(65.83,60.00)
\bezier{14}(34.17,60.00)(34.17,56.83)(42.09,55.03)
\bezier{20}(42.09,55.03)(50.00,53.61)(57.91,55.03)
\bezier{14}(57.91,55.03)(65.83,56.83)(65.83,60.00)
\bezier{264}(24.00,35.00)(50.00,55.00)(76.00,35.00)
\bezier{56}(39.30,55.79)(46.32,39.30)(56.32,20.53)
\bezier{56}(60.70,55.79)(53.68,39.30)(43.68,20.53)
\put(50.00,31.00){\circle*{0.00}}
\put(49.00,30.00){\circle*{0.00}}
\put(50.00,29.00){\circle*{0.00}}
\put(49.00,28.00){\circle*{0.00}}
\put(50.00,27.00){\circle*{0.00}}
\put(48.00,27.00){\circle*{0.00}}
\put(49.00,26.00){\circle*{0.00}}
\put(47.00,26.00){\circle*{0.00}}
\put(48.00,25.00){\circle*{0.00}}
\put(50.00,25.00){\circle*{0.00}}
\put(49.00,24.00){\circle*{0.00}}
\put(47.00,24.00){\circle*{0.00}}
\put(46.00,23.00){\circle*{0.00}}
\put(48.00,23.00){\circle*{0.00}}
\put(50.00,23.00){\circle*{0.00}}
\put(49.00,22.00){\circle*{0.00}}
\put(47.00,22.00){\circle*{0.00}}
\put(45.00,22.00){\circle*{0.00}}
\put(46.00,21.00){\circle*{0.00}}
\put(48.00,21.00){\circle*{0.00}}
\put(50.00,21.00){\circle*{0.00}}
\put(44.00,21.00){\circle*{0.00}}
\put(47.02,20.18){\circle*{0.00}}
\put(48.95,20.18){\circle*{0.00}}
\put(51.00,30.00){\circle*{0.00}}
\put(52.00,29.00){\circle*{0.00}}
\put(51.00,28.00){\circle*{0.00}}
\put(52.00,27.00){\circle*{0.00}}
\put(51.00,26.00){\circle*{0.00}}
\put(53.00,26.00){\circle*{0.00}}
\put(52.00,25.00){\circle*{0.00}}
\put(54.00,25.00){\circle*{0.00}}
\put(51.00,24.00){\circle*{0.00}}
\put(53.00,24.00){\circle*{0.00}}
\put(54.00,23.00){\circle*{0.00}}
\put(52.00,23.00){\circle*{0.00}}
\put(51.00,22.00){\circle*{0.00}}
\put(53.00,22.00){\circle*{0.00}}
\put(55.00,22.00){\circle*{0.00}}
\put(54.00,21.00){\circle*{0.00}}
\put(52.00,21.00){\circle*{0.00}}
\put(56.00,21.00){\circle*{0.00}}
\put(52.98,20.18){\circle*{0.00}}
\put(51.05,20.18){\circle*{0.00}}
\put(54.00,68.00){\vector(-1,-2){3.50}}
\put(55.00,68.00){\makebox(0,0)[lc]{C (Center of projection)}}
\put(29.00,65.00){\vector(1,-1){4.50}}
\put(28.50,65.00){\makebox(0,0)[rc]{Conformal infinity}}
\put(20.00,40.00){\vector(1,0){5.0}}
\put(19.50,40.00){\makebox(0,0)[rc]{P}}  
\bezier{36}(23.86,35.09)(21.75,31.75)(24.74,28.25)
\bezier{92}(38.25,22.63)(49.82,20.88)(61.40,22.63)
\bezier{60}(24.91,28.07)(28.77,24.39)(38.42,22.63)
\bezier{36}(76.14,35.09)(78.25,31.75)(75.26,28.25)
\bezier{60}(75.09,28.07)(71.23,24.39)(61.58,22.63)
\put(50.00,32.63){\line(0,-1){14.56}}
\put(15.96,31.75){\line(5,-2){4.91}}
\put(24.74,28.25){\line(5,-2){13.68}}
\put(44.04,20.53){\line(5,-2){5.96}}
\put(15.96,31.75){\line(5,2){4.56}}
\bezier{30}(25.61,39.65)(33.16,47.54)(39.30,55.80)
\bezier{26}(34.04,60.00)(34.04,50.35)(25.09,39.30)
\put(25.79,39.82){\line(-6,-5){9.82}}
\put(71.93,47.37){\vector(-3,-2){9.65}}
\put(72.28,47.37){\makebox(0,0)[lc]{Initial surface}}
\put(84.04,31.75){\line(-5,-2){4.91}}
\put(75.26,28.25){\line(-5,-2){13.68}}
\put(84.04,31.75){\line(-5,2){4.56}}
\bezier{30}(74.39,39.65)(66.84,47.54)(60.70,55.80)
\bezier{26}(65.96,60.00)(65.96,50.35)(74.91,39.30)
\put(74.21,39.82){\line(6,-5){9.82}}
\put(55.96,20.53){\line(-5,-2){5.96}}
\thicklines
\put(20.88,29.79){\line(5,-2){3.86}}
\put(38.42,22.77){\line(5,-2){5.61}}
\bezier{16}(20.53,33.68)(19.47,31.75)(20.53,29.82)
\bezier{36}(44.00,20.50)(50.00,19.65)(56.00,20.50)
\put(44.04,20.70){\line(1,2){5.88}}
\put(49.91,32.46){\line(0,1){27.54}}
\bezier{16}(20.53,33.51)(22.11,34.39)(23.68,35.09)
\put(20.53,33.51){\line(5,6){5.12}}
\put(50.00,60.00){\line(-6,-5){24.21}}
\put(20.53,29.87){\line(1,2){4.87}}
\put(79.12,29.79){\line(-5,-2){3.86}}
\put(61.58,22.77){\line(-5,-2){5.61}}
\bezier{16}(79.47,33.68)(80.53,31.75)(79.47,29.82)
\put(55.96,20.70){\line(-1,2){5.88}}
\bezier{16}(79.47,33.51)(77.89,34.39)(76.32,35.09)
\put(79.47,33.51){\line(-5,6){5.12}}
\put(50.00,60.00){\line(6,-5){24.21}}
\put(79.47,29.87){\line(-1,2){4.87}}
\bezier{80}(75.00,28.00)(65.67,32.33)(62.00,23.00)
\bezier{80}(25.00,28.00)(34.33,32.33)(38.00,23.00)
\end{picture}
\caption{The identification surfaces near the collapse point C in stereographic 
coordinates. AdS spacetime is the interior of the lightly dotted hyperboloid. 
The hyperboloid itself represents conformal infinity. The initial surface is 
a Minkowski hyperboloid (like that of Fig.~5) and in that sense is shown in 
its true metric. The triangular regions on the infinity hyperboloid, 
one of which is dotted, are the part of \scri\ that can be shown in this
coordinate neighborhood} 
\end{figure}

The tent has a simple, pyramid shape in a stereographic mapping centered 
at C. Since all geodesics through the center of the map are represented 
by straight lines in such a map, the sides of the tent are timelike planes
(that is, linear spaces in stereographic coordinates), and the folds are 
spacelike straight lines. Figure~13 shows a tent with no 
corners but four folds. This can be the 
spacetime fundamental domain for the $k=3,\,g=0$ three-black-hole of 
Fig.~10 or for the $k=1,\,g=1$ toroidal black hole of Fig.~12, depending 
on the identification rule.
In the three-black-hole case two of the folds, on opposite sides, are fixed
points of the identifications $a$ and $b$ of Fig.~10 that generate the 
group. The other two folds are fixed points of $ab$ and of $ba$. For the 
toroidal black hole the fixed points of $a$ and of $b$ of Fig.~12 are not 
folds, they would be horizontal lines through the tip of the tent. 
Instead the folds are fixed points of $aba^{-1}b^{-1}$ and its three 
cyclic permutations. In each 
case the folds are fixed points of transformations associated with a horizon.
All the other fixed points lie outside of the fundamental domain. 

Because the stereographic picture is centered at a particular time, it 
can be misleading in that it does not exhibit the 
time symmetry about the initial surface, nor the early history before 
the time-symmetric moment. The time-independent sausage coordinates 
are more suitable for the global view of a black hole spacetime. 
Since the BTZ black hole (Fig.~14a) involves 
only one identification, its fundamental domain has only one geodesic of 
fixed points to the future of the initial surface, 
the $r=0$ line in Fig.~2a. The sides of the tent are the surfaces 
$\phi=0$ and $\phi=2\pi$. Fig.~14b is the sausage coordinate version of 
Fig.~13.

\begin{figure}
\unitlength 0.90mm
\linethickness{0.4pt}
\begin{picture}(121.81,77.01)(0,0)
\bezier{80}(42.21,16.71)(42.21,13.47)(35.11,11.84)
\bezier{80}(35.11,11.84)(28.00,10.22)(20.89,11.84)
\bezier{80}(20.89,11.84)(13.79,13.47)(13.79,16.71)
\bezier{80}(42.21,44.56)(42.21,41.88)(35.11,40.55)
\bezier{80}(35.11,40.55)(28.00,39.22)(20.89,40.55)
\bezier{80}(20.89,40.55)(13.79,41.88)(13.79,44.56)
\bezier{32}(32.26,40.16)(31.40,40.97)(33.69,41.78)
\bezier{44}(33.69,41.78)(36.61,42.47)(39.37,41.78)
\bezier{32}(23.74,48.95)(24.60,48.14)(22.31,47.34)
\bezier{44}(22.31,47.34)(19.39,46.64)(16.63,47.34)
\bezier{80}(42.21,72.40)(42.21,74.71)(35.11,75.86)
\bezier{80}(35.11,75.86)(28.00,77.01)(20.89,75.86)
\bezier{80}(20.89,75.86)(13.79,74.71)(13.79,72.40)
\bezier{80}(42.21,72.40)(42.21,70.09)(35.11,68.94)
\bezier{80}(35.11,68.94)(28.00,67.78)(20.89,68.94)
\bezier{80}(20.89,68.94)(13.79,70.09)(13.79,72.40)
\put(13.79,16.71){\line(0,1){55.68}}
\put(42.21,16.71){\line(0,1){55.68}}
\put(28.00,5.58){\makebox(0,0)[cc]{({\bf a})}}
\thicklines
\multiput(16.98,69.89)(0.46,0.12){46}{\line(1,0){0.46}}
\multiput(16.63,13.37)(0.38,0.12){58}{\line(1,0){0.38}}
\bezier{536}(16.98,69.89)(47.90,38.66)(16.63,13.37)
\bezier{148}(37.95,75.18)(42.92,67.01)(42.21,55.69)
\bezier{128}(40.79,33.98)(37.95,41.96)(40.79,50.13)
\bezier{80}(40.79,50.13)(42.21,55.69)(42.21,55.69)
\bezier{88}(40.79,33.98)(42.21,29.52)(42.21,27.85)
\bezier{44}(42.21,27.85)(42.21,23.40)(38.66,20.33)
\bezier{34}(37.95,75.18)(29.90,67.57)(25.87,59.03)
\bezier{228}(25.87,59.03)(19.47,45.30)(30.13,30.64)
\bezier{23}(30.13,30.64)(32.97,25.81)(38.66,20.33)
\bezier{148}(16.63,13.37)(13.08,22.10)(13.79,33.42)
\bezier{128}(15.21,56.25)(18.05,47.16)(15.21,38.99)
\bezier{80}(15.21,38.99)(13.79,33.42)(13.79,33.42)
\bezier{88}(15.21,56.25)(13.79,59.59)(13.79,62.38)
\bezier{44}(13.79,62.93)(13.79,65.72)(16.98,69.89)
\thinlines
\put(15.92,69.43){\vector(4,1){0.2}}
\multiput(8.10,67.39)(0.43,0.11){18}{\line(1,0){0.43}}
\put(7.39,67.94){\makebox(0,0)[cc]{P}}
\put(42.92,44.56){\makebox(0,0)[lc]{$t=0$}}
\bezier{40}(42.21,16.71)(42.21,18.66)(38.66,20.33)
\bezier{19}(38.66,20.33)(31.55,22.84)(25.87,22.28)
\bezier{17}(25.87,22.28)(18.41,21.73)(14.85,18.94)
\bezier{20}(14.85,18.94)(13.79,18.00)(13.79,16.71)
\bezier{32}(13.79,44.56)(13.50,46.12)(16.63,47.38)
\bezier{10}(16.63,47.38)(19.85,48.64)(23.45,49.01)
\bezier{40}(23.45,49.01)(27.05,49.38)(30.75,49.16)
\bezier{14}(30.75,49.16)(36.24,48.72)(39.94,47.08)
\bezier{24}(39.94,47.08)(41.84,46.41)(42.21,44.56)
\bezier{80}(109.73,16.71)(109.73,19.96)(102.62,21.59)
\bezier{28}(88.41,21.59)(90.90,21.95)(93.03,22.00)
\bezier{28}(102.62,21.59)(100.13,21.95)(98.00,22.00)
\bezier{80}(88.41,21.59)(81.30,19.96)(81.30,16.71)
\bezier{80}(109.73,16.71)(109.73,13.47)(102.62,11.84)
\bezier{80}(102.62,11.84)(95.51,10.22)(88.41,11.84)
\bezier{80}(88.41,11.84)(81.30,13.47)(81.30,16.71)
\bezier{20}(109.73,44.56)(109.73,47.23)(102.62,48.57)
\bezier{20}(102.62,48.57)(95.51,49.90)(88.41,48.57)
\bezier{20}(88.41,48.57)(81.30,47.23)(81.30,44.56)
\bezier{80}(109.73,44.56)(109.73,41.88)(102.62,40.55)
\bezier{80}(102.62,40.55)(95.51,39.22)(88.41,40.55)
\bezier{80}(88.41,40.55)(81.30,41.88)(81.30,44.56)
\bezier{32}(82.01,43.17)(84.50,43.45)(86.99,42.71)
\bezier{44}(86.99,42.71)(89.12,41.76)(86.99,40.85)
\bezier{32}(99.78,40.16)(99.78,41.49)(101.20,41.78)
\bezier{44}(101.20,41.78)(104.12,42.47)(106.89,41.78)
\bezier{36}(91.20,49.01)(92.35,47.68)(88.27,47.08)
\bezier{6}(88.27,47.08)(86.47,46.78)(84.00,47.38)
\bezier{12}(109.02,45.95)(108.26,45.84)(107.31,45.87)
\bezier{15}(107.31,45.87)(100.58,46.63)(104.19,48.27)
\bezier{80}(109.73,72.40)(109.73,74.71)(102.62,75.86)
\bezier{80}(102.62,75.86)(95.51,77.01)(88.41,75.86)
\bezier{80}(88.41,75.86)(81.30,74.71)(81.30,72.40)
\bezier{80}(109.73,72.40)(109.73,70.09)(102.62,68.94)
\bezier{80}(102.62,68.94)(95.51,67.78)(88.41,68.94)
\bezier{80}(88.41,68.94)(81.30,70.09)(81.30,72.40)
\put(81.30,16.71){\line(0,1){55.68}}
\put(109.73,16.71){\line(0,1){55.68}}
\thicklines
\bezier{252}(93.38,25.62)(80.59,42.51)(93.38,54.02)
\bezier{248}(93.38,54.02)(106.17,40.29)(93.38,25.62)
\bezier{208}(109.02,29.52)(104.54,40.84)(109.02,57.92)
\bezier{36}(109.02,57.92)(109.73,54.86)(109.73,52.35)
\bezier{116}(109.73,52.35)(108.31,46.51)(109.73,36.20)
\bezier{52}(109.73,36.20)(109.73,32.86)(109.02,29.52)
\bezier{96}(81.30,51.80)(82.72,42.89)(81.30,38.43)
\bezier{52}(82.01,59.03)(81.30,54.86)(81.30,51.80)
\bezier{48}(82.01,31.75)(81.30,36.20)(81.30,38.43)
\thinlines
\bezier{50}(82.01,59.03)(86.77,46.97)(82.01,31.75)
\bezier{16}(97.65,35.65)(99.54,36.94)(99.78,37.87)
\bezier{54}(99.78,37.87)(109.49,46.97)(97.65,62.38)
\bezier{108}(97.65,35.65)(90.94,42.39)(91.25,49.01)
\bezier{28}(91.25,49.01)(90.94,49.07)(91.96,52.91)
\bezier{20}(91.96,52.91)(93.30,56.75)(97.65,62.93)
\bezier{44}(97.65,35.65)(96.94,34.26)(96.23,29.52)
\bezier{23}(96.23,29.52)(95.87,23.68)(95.51,16.71)
\thicklines
\bezier{152}(82.01,31.75)(91.25,29.15)(95.51,16.71)
\bezier{156}(95.51,16.71)(99.07,29.52)(109.02,29.52)
\bezier{84}(93.38,25.62)(95.28,26.74)(95.51,16.71)
\bezier{148}(82.01,59.03)(92.67,61.26)(95.51,72.40)
\bezier{148}(95.51,72.40)(99.78,60.71)(109.02,57.92)
\bezier{132}(93.38,54.02)(95.28,60.33)(95.51,72.40)
\thinlines
\bezier{19}(95.51,72.40)(96.94,62.56)(97.65,62.93)
\put(92.67,54.16){\vector(4,-1){0.2}}
\multiput(70.64,58.48)(0.61,-0.12){36}{\line(1,0){0.61}}
\put(70.64,59.03){\vector(1,0){11.02}}
\put(69.93,59.03){\makebox(0,0)[rc]{P}}
\put(121.10,57.92){\vector(-1,0){11.73}}
\bezier{76}(121.10,59.03)(115.41,62.38)(109.73,63.49)
\put(97.65,63.49){\vector(-4,-1){0.2}}
\bezier{68}(109.73,63.49)(104.52,64.60)(97.65,63.49)
\put(121.81,58.48){\makebox(0,0)[lc]{P}}
\put(95.51,5.58){\makebox(0,0)[cc]{({\bf b})}}
\bezier{130}(88.53,47.62)(87.34,39.27)(89.24,32.40)
\end{picture}
\caption{Fundamental regions and their boundary ``tents" in sausage 
coordinates, for ({\bf a}) the BTZ black hole and ({\bf b}) a 
three-black-hole or toroidal black hole configuration}
\end{figure}
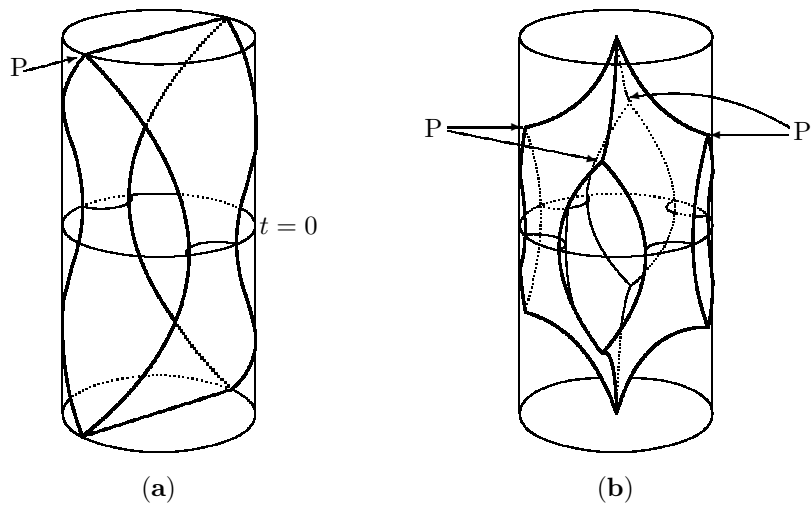

Since folds are spacelike they extend to infinity, and therefore the 
initial fundamental domain must also have asymptotic regions. Conversely,
the tent of an initial state without asymptotic regions has only corners 
and a tip but no folds: any closed time-symmetric AdS universe always 
collapses to a point in the finite time $\pi\ell/2$.

The holes in the tent are important for the black hole interpretation, for 
they are the regions at infinity, \scri. The edges of the holes of course 
disappear once the identifications are made, and the only remaining 
boundaries of \scri\ appear as points such as those marked P in 
the figure. The backwards lightcone from P is the boundary of the past of 
\scri, i.e.\ the horizon. It surrounds the singularity whose end is P. 
It is now clear that all the initial configurations that have a horizon 
in the sense of section 2.4 do have spacetime horizons and hence are black 
holes: a horizon word extended to spacetime is an identification that has
fixed points along some fold of the tent-shaped boundary of the spacetime 
fundamental domain. The intersection of the fold with conformal infinity
is an endpoint of a \scri, and the backwards 
lightcone of that endpoint is the spacetime horizon.

For a given fold we can consider a region in the fundamental domain 
sufficiently near infinity (spatially) and the fold (temporally) so that 
the only relevant identification is the one that has fixed points on that 
fold (because the other identifications would move points out of the 
region). In that region the spacetime is then indistinguishable from that 
of a BTZ black hole, and the spacetime horizon behaves in the same 
way as a BTZ black hole horizon. For example, the backward lightcone 
from P does intersect the initial surface in the minimal horizon 
geodesic. As we follow the horizon further backward in time it changes from the BTZ 
behavior only when it encounters other horizons or another part of 
itself, coming from another copy of the point P in the fundamental domain.
For example, in the toroidal black hole interpretation of Fig.~14, all 
four openings of the tent are parts of one \scri, and there is a single 
spacetime horizon consisting of the four ``quarter" backwards lightcones 
from the four copies of the point P. As we go backwards in time below 
the initial surface these lightcones eventually touch and merge and shown 
in Fig.~15.

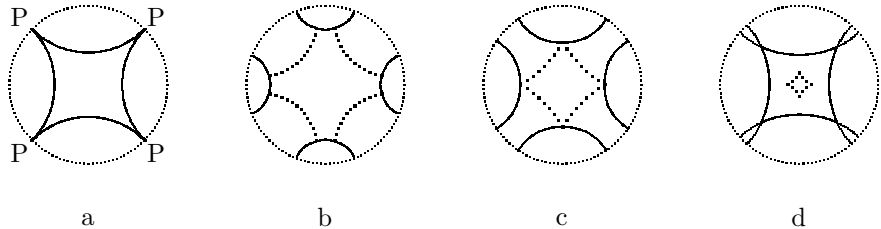
\begin{figure}
\unitlength 1.05mm
\linethickness{0.4pt}
\begin{picture}(120.00,30.00)(9,0)
\bezier{15}(10.00,20.00)(10.00,24.14)(12.93,27.07)
\bezier{15}(12.93,27.07)(15.86,30.00)(20.00,30.00)
\bezier{15}(30.00,20.00)(30.00,24.14)(27.07,27.07)
\bezier{15}(27.07,27.07)(24.14,30.00)(20.00,30.00)
\bezier{15}(10.00,20.00)(10.00,15.86)(12.93,12.93)
\bezier{15}(12.93,12.93)(15.86,10.00)(20.00,10.00)
\bezier{15}(30.00,20.00)(30.00,15.86)(27.07,12.93)
\bezier{15}(27.07,12.93)(24.14,10.00)(20.00,10.00)
\bezier{15}(40.00,20.00)(40.00,24.14)(42.93,27.07)
\bezier{15}(70.00,20.00)(70.00,24.14)(72.93,27.07)
\bezier{15}(100.00,20.00)(100.00,24.14)(102.93,27.07)
\bezier{15}(42.93,27.07)(45.86,30.00)(50.00,30.00)
\bezier{15}(72.93,27.07)(75.86,30.00)(80.00,30.00)
\bezier{15}(102.93,27.07)(105.86,30.00)(110.00,30.00)
\bezier{15}(60.00,20.00)(60.00,24.14)(57.07,27.07)
\bezier{15}(90.00,20.00)(90.00,24.14)(87.07,27.07)
\bezier{15}(120.00,20.00)(120.00,24.14)(117.07,27.07)
\bezier{15}(57.07,27.07)(54.14,30.00)(50.00,30.00)
\bezier{15}(87.07,27.07)(84.14,30.00)(80.00,30.00)
\bezier{15}(117.07,27.07)(114.14,30.00)(110.00,30.00)
\bezier{15}(40.00,20.00)(40.00,15.86)(42.93,12.93)
\bezier{15}(70.00,20.00)(70.00,15.86)(72.93,12.93)
\bezier{15}(100.00,20.00)(100.00,15.86)(102.93,12.93)
\bezier{15}(42.93,12.93)(45.86,10.00)(50.00,10.00)
\bezier{15}(72.93,12.93)(75.86,10.00)(80.00,10.00)
\bezier{15}(102.93,12.93)(105.86,10.00)(110.00,10.00)
\bezier{15}(60.00,20.00)(60.00,15.86)(57.07,12.93)
\bezier{15}(90.00,20.00)(90.00,15.86)(87.07,12.93)
\bezier{15}(120.00,20.00)(120.00,15.86)(117.07,12.93)
\bezier{15}(57.07,12.93)(54.14,10.00)(50.00,10.00)
\bezier{15}(87.07,12.93)(84.14,10.00)(80.00,10.00)
\bezier{15}(117.07,12.93)(114.14,10.00)(110.00,10.00)
\bezier{30}(46.38,29.13)(47.25,27.25)(50.00,26.96)
\bezier{30}(53.62,29.13)(52.75,27.25)(50.00,26.96)
\bezier{30}(40.87,23.62)(42.75,22.75)(43.04,20.00)
\bezier{30}(40.87,16.38)(42.75,17.25)(43.04,20.00)
\bezier{50}(12.78,12.93)(15.71,15.86)(15.71,20.00)
\bezier{50}(12.93,27.03)(15.86,24.10)(20.00,24.10)
\bezier{50}(27.07,27.03)(24.14,24.10)(20.00,24.10)
\bezier{50}(27.22,12.93)(24.29,15.86)(24.29,20.00)
\bezier{50}(27.22,27.07)(24.29,24.14)(24.29,20.00)
\bezier{50}(27.07,12.97)(24.14,15.90)(20.00,15.90)
\bezier{50}(12.93,12.97)(15.86,15.90)(20.00,15.90)
\bezier{50}(12.78,27.07)(15.71,24.14)(15.71,20.00)
\thicklines
\bezier{12}(42.90,21.16)(47.83,22.17)(48.84,27.10)
\bezier{12}(42.90,18.84)(47.83,17.83)(48.84,12.90)
\bezier{12}(57.10,21.16)(52.17,22.17)(51.16,27.10)
\bezier{12}(57.10,18.84)(52.17,17.83)(51.16,12.90)
\thinlines
\bezier{30}(46.38,10.87)(47.25,12.75)(50.00,13.04)
\bezier{30}(53.62,10.87)(52.75,12.75)(50.00,13.04)
\bezier{30}(59.13,23.62)(57.25,22.75)(56.96,20.00)
\bezier{30}(59.13,16.38)(57.25,17.25)(56.96,20.00)
\bezier{40}(74.44,28.33)(76.33,25.33)(80.00,25.33)
\bezier{40}(85.56,28.33)(83.67,25.33)(80.00,25.33)
\bezier{40}(88.33,14.44)(85.33,16.33)(85.33,20.00)
\bezier{40}(88.33,25.56)(85.33,23.67)(85.33,20.00)
\bezier{40}(71.67,25.56)(74.67,23.67)(74.67,20.00)
\bezier{40}(71.67,14.44)(74.67,16.33)(74.67,20.00)
\bezier{40}(85.56,11.67)(83.67,14.67)(80.00,14.67)
\bezier{40}(74.44,11.67)(76.33,14.67)(80.00,14.67)
\thicklines
\bezier{8}(80.00,25.33)(82.00,22.00)(85.33,20.00)
\bezier{8}(80.00,25.33)(78.00,22.00)(74.67,20.00)
\bezier{8}(80.00,14.67)(82.00,18.00)(85.33,20.00)
\bezier{8}(80.00,14.67)(78.00,18.00)(74.67,20.00)
\thinlines
\bezier{30}(106.22,20.00)(106.22,22.50)(105.22,24.78)
\bezier{10}(105.22,24.78)(104.56,26.20)(103.56,27.30)
\bezier{30}(106.22,20.00)(106.22,17.50)(105.22,15.22)
\bezier{10}(105.22,15.22)(104.56,13.80)(103.56,12.70)
\bezier{30}(110.00,16.22)(107.50,16.22)(105.22,15.22)
\bezier{10}(105.22,15.22)(103.80,14.56)(102.70,13.56)
\bezier{30}(110.00,16.22)(112.50,16.22)(114.78,15.22)
\bezier{10}(114.78,15.22)(116.20,14.56)(117.30,13.56)
\bezier{30}(113.78,20.00)(113.78,17.50)(114.78,15.22)
\bezier{30}(113.78,20.00)(113.78,22.50)(114.78,24.78)
\bezier{30}(110.00,23.78)(112.50,23.78)(114.78,24.78)
\bezier{30}(110.00,23.78)(107.50,23.78)(105.22,24.78)
\bezier{10}(114.78,15.22)(115.44,13.80)(116.44,12.70)
\bezier{10}(114.78,24.78)(115.44,26.20)(116.44,27.30)
\bezier{10}(114.78,24.78)(116.20,25.44)(117.30,26.44)
\bezier{10}(105.22,24.78)(103.80,25.44)(102.70,26.44)
\thicklines
\bezier{3}(110.00,21.50)(109.50,20.50)(108.50,20.00)
\bezier{3}(110.00,21.50)(110.50,20.50)(111.50,20.00)
\bezier{3}(110.00,18.50)(109.50,19.50)(108.50,20.00)
\bezier{3}(110.00,18.50)(110.50,19.50)(111.50,20.00)
\thinlines
\put(27.50,27.50){\makebox(0,0)[lb]{P}}
\put(12.50,27.50){\makebox(0,0)[rb]{P}}
\put(12.50,12.50){\makebox(0,0)[rt]{P}}
\put(27.50,12.50){\makebox(0,0)[lt]{P}}
\put(110.00,3.30){\makebox(0,0)[cc]{d}}
\put(20.00,3.00){\makebox(0,0)[cc]{a}}
\put(50.00,3.30){\makebox(0,0)[cc]{b}}
\put(80.00,3.00){\makebox(0,0)[cc]{c}}
\end{picture}
\caption{Slices of the sausage in Fig.~14 to show the time development 
of the horizon. Part a is the latest and part d the earliest sausage time.
The geometry of each time slice is the constant curvature space 
represented by a Poincar\'e disk. The geodesics shown by solid lines are 
to be identified as before for the toroidal black hole. Where these 
geodesics intersect we have a fixed point of some identification, a 
physical singularity. Slice a is at the sausage time of the 
end point P of \scri. As we go backwards in time, the horizon 
(dotted arcs of circles) spreads out from those points at infinity. 
Slice b is the moment of time-symmetry. The horizon remains smooth 
until slice c, when its different parts meet 
each other at the identification surfaces. Prior to that time the event 
horizon has four kinks} 
\end{figure}

\subsection{Fixed Points at Infinity}

As the above examples of multi-black-hole time developments show,
any time-symmetric initial state with an asymptotic region ending at a 
horizon is isometric to a corresponding region of 
a BTZ black hole, so that each such region will look like a black hole 
from infinity for at least a finite time. It is maybe not so clear 
whether this is also true for the unlimited time necessary for a true black 
hole, for example because other singularities (fixed points) might intervene. 
By an interesting method due to {\AA}minneborg, Bengtsson and Holst 
\cite{ABH} one can directly find all of 
the universal covering space of \scri\ from a knowledge of spatial infinity 
on an initial surface. (The universal covering space gives information 
about horizons and is natural in many contexts, for example 
topological censorship questions reduce to existence of certain geodesics 
in AdS space \cite{DBS}.)

Since our black holes are quotient spaces of AdS space, the covering 
space of their \scri\ will be a subset of conformal infinity of AdS space.
To describe this conformal infinity in a finite way we follow the 
usual Penrose procedure and multiply 
the AdS metric by a factor so that the resulting metric is finite in the 
asymptotic region. An obviously suitable 
conformal factor in Eq (\ref{sausge}) is $(1-(\rho/\ell)^2)^2$, giving the 
metric at infinity, $\rho=\ell$,
$$ds^2_\infty = 4 (dt^2 + \ell^2 d\theta^2)$$
This is the flat metric of a cylinder of radius $\ell$. 

Consider first the covering space of \scri\ of a single black hole in 
this description, and recall that the identification is a ``Lorentz 
boost" in the 
embedding space (\ref{ads}). As we apply this transformation $n$ times to 
get the $n$th tile of the covering space, we are boosting the fundamental 
domain to the limiting velocity, and since the identification boundaries 
are timelike in our description, they become two null surfaces in the limit
$n\rightarrow\pm\infty$. These null surfaces (called ``singularity 
surfaces" in \cite{HP}) are then of 
course invariant under the identification transformation. Hence, if 
${\bf K}=\partial/\partial\phi$ is the Killing vector corresponding to 
the identification, these surfaces are described by ${\bf K}^2=0$. They 
intersect where the vector {\bf K} itself vanishes, that is at the fixed 
points at infinity on 
the initial surface and at the singularity inside the black hole.

The intersection of these surfaces with 
conformal infinity of AdS space is the boundary ($n\rightarrow\infty$) 
of the covering space of \scri. To find it we only need to draw 
null lines from the endpoints of the horizon at $t=0$ toward each other 
(Fig.~16a). Their future
intersection is the nearest future fixed point to this initial 
surface, it is the end of \scri, and the covering space of \scri\ is the 
diamond-shaped region between the future and past null lines. Furthermore 
the future null lines are also the intersection of the covering space of the
horizon with conformal 
infinity, since the horizon is the backward null cone from the end of \scri.
Thus a knowledge of the initial fixed points gives us the ``holographic" 
information about the exterior and the horizon of the black hole.

\begin{figure}
\unitlength 1.20mm
\linethickness{0.4pt}
\begin{picture}(106.00,32.00)(10,0)
\put(10.00,20.00){\line(1,-1){10.00}}
\put(20.00,10.00){\line(1,1){20.00}}
\put(40.00,30.00){\line(1,-1){10.00}}
\put(50.00,20.00){\line(-1,-1){10.00}}
\put(40.00,10.00){\line(-1,1){20.00}}
\put(20.00,30.00){\line(-1,-1){10.00}}
\put(10.00,8.00){\line(0,1){24.00}}
\put(50.00,32.00){\line(0,-1){24.00}}
\thicklines
\bezier{104}(40.00,30.00)(48.00,20.00)(40.00,10.00)
\thinlines
\bezier{72}(40.00,30.00)(47.00,22.33)(47.00,20.00)
\bezier{76}(40.00,30.00)(48.67,21.00)(48.67,20.00)
\bezier{72}(40.00,10.00)(47.00,17.67)(47.00,20.00)
\bezier{76}(40.00,10.00)(48.67,19.00)(48.67,20.00)
\thicklines
\bezier{104}(20.00,30.00)(12.00,20.00)(20.00,10.00)
\thinlines
\bezier{72}(20.00,30.00)(13.00,22.33)(13.00,20.00)
\bezier{76}(20.00,30.00)(11.33,21.00)(11.33,20.00)
\bezier{72}(20.00,10.00)(13.00,17.67)(13.00,20.00)
\bezier{76}(20.00,10.00)(11.33,19.00)(11.33,20.00)
\thicklines
\put(40.00,10.00){\vector(0,1){20.00}}
\put(20.00,30.00){\line(0,-1){20.00}}
\thinlines
\put(30.00,20.00){\vector(1,0){20.00}}
\put(30.00,20.00){\vector(-1,0){20.00}}
\put(51.00,20.00){\makebox(0,0)[lc]{$\phi$}}
\put(65.00,20.00){\line(1,-1){10.00}}
\put(75.00,10.00){\line(1,1){20.00}}
\put(95.00,30.00){\line(1,-1){10.00}}
\put(105.00,20.00){\line(-1,-1){10.00}}
\put(95.00,10.00){\line(-1,1){20.00}}
\put(75.00,30.00){\line(-1,-1){10.00}}
\put(65.00,8.00){\line(0,1){24.00}}
\put(105.00,32.00){\line(0,-1){24.00}}
\bezier{84}(85.00,20.00)(81.79,17.95)(75.00,20.00)
\bezier{84}(85.00,20.00)(88.21,22.05)(95.00,20.00)
\put(65.00,20.00){\vector(-2,-1){0.2}}
\bezier{84}(75.00,20.00)(68.21,22.05)(65.00,20.00)
\put(105.00,20.00){\vector(2,1){0.2}}
\bezier{84}(95.00,20.00)(102.05,17.95)(105.00,20.00)
\put(106.00,20.00){\makebox(0,0)[lc]{$\varphi$}}
\thicklines
\bezier{64}(95.00,10.00)(92.95,13.21)(95.00,20.00)
\put(95.00,30.00){\vector(-1,2){0.2}}
\bezier{84}(95.00,20.00)(96.92,26.67)(95.00,30.00)
\thinlines
\put(95.00,30.51){\makebox(0,0)[cb]{$\tau$}}
\put(40.00,30.51){\makebox(0,0)[cb]{$t$}}
\thicklines
\bezier{84}(75.00,10.00)(72.95,13.21)(75.00,20.00)
\bezier{84}(75.00,20.00)(76.92,26.67)(75.00,30.00)
\bezier{60}(95.00,10.00)(99.00,17.00)(98.83,19.00)
\bezier{68}(98.83,19.00)(98.00,25.17)(95.00,30.00)
\thinlines
\bezier{88}(95.00,10.00)(101.50,17.50)(101.50,19.00)
\bezier{86}(101.50,19.00)(101.50,20.67)(95.00,30.17)
\bezier{86}(95.00,10.00)(103.50,18.83)(102.83,19.00)
\bezier{84}(102.83,19.00)(104.50,19.50)(95.00,30.00)
\thicklines
\bezier{60}(75.00,30.00)(71.00,23.00)(71.17,21.00)
\bezier{68}(71.17,21.00)(72.00,14.83)(75.00,10.00)
\thinlines
\bezier{88}(75.00,30.00)(68.50,22.50)(68.50,21.00)
\bezier{86}(68.50,21.00)(68.50,19.33)(75.00,9.83)
\bezier{86}(75.00,30.00)(66.50,21.17)(67.17,21.00)
\bezier{84}(67.17,21.00)(65.50,20.50)(75.00,10.00)
\put(85.00,3.00){\makebox(0,0)[cc]{({\bf b})}}
\put(30.00,3.00){\makebox(0,0)[cc]{({\bf a})}}
\end{picture}
\caption{Universal covering spaces of conformal infinity \scri\ for 
({\bf a}) non-rotating, ({\bf b})~rotating black holes on the conformal 
infinity cylinder of AdS space. To show the cylinder in this 
flat picture it has been cut along the vertical lines, which are 
to be identified with each other in each part of the figure. The
angle $\phi$ resp.\ $\varphi$ runs from $-\infty$ to $+\infty$, in
the direction of the arrow, in each of the diamond-shaped regions. The 
intersection of the fundamental domain with conformal infinity of AdS 
space is shown by the heavier boundaries. These 
boundaries are at the values $0$ and $2\pi$ of $\phi$ resp.\ $\varphi$. A 
few of the tiles obtained by the isometries that change these angles by 
$2\pi n$ are shown for positive multiples $n$. In the limit 
$n\rightarrow\pm\infty$ the tiles converge to the null boundaries of the two 
diamond-shaped regions. These null boundaries intersect on the initial 
surface ($t=0$ resp.\ $\tau=0$) at the $n\rightarrow\infty$ limit at 
conformal infinity of the initial surface, and they end in the future at the 
end of \scri} 
\end{figure}
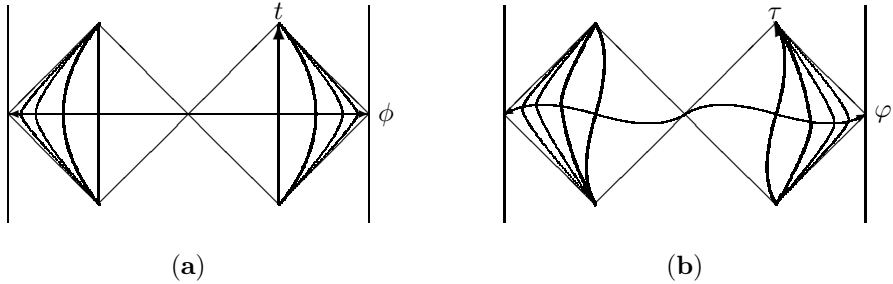

The situation for time-symmetric multi-black-holes is similar, except 
that the construction yields an infinite number of copies of the 
covering space of \scri. We saw in Fig.~10 that each horizon word has two 
fixed points at spatial infinity. Each such pair yields a diamond-shaped 
region for which the transformation of its horizon word looks like 
that of Fig.~16a, and which is free of fixed points. 

\section{Angular Momentum}

In the metric (\ref{BTZ}) for the static BTZ black hole, introduce new 
coordinates $T,\,\varphi,\,R$, 
\begin{eqnarray}
t = T + \left({J\over 2m}\right)\varphi\nonumber \\ 
\phi = \varphi + \left({J\over 2m\ell^2}\right)T \\ 
R^2=r^2\left(1-{J^2\over 4m^2\ell^2}\right) + {J^2\over 4m}\nonumber
\label{coord}
\end{eqnarray}
where $J<2m\ell$ is a constant with dimension of length, and define another new 
constant 
\begin{equation}
M = m + {J^2\over 4m\ell^2}.
\label{mass}
\end{equation}
In terms of these new quantities the metric (\ref{BTZ}) may be written as
\begin{equation}
ds^2 = -N^2 dT^2 + N^{-2}dR^2 + R^2\left(d\varphi + {J\over 2R^2} 
dT\right)^2
\label{BTZJ}
\end{equation}
where
$$N^2 = \left({R\over\ell}\right)^2 - M + \left({J\over 2R}\right)^2.$$

The metric (\ref{BTZJ}) now looks like a (2+1)-dimensional analog of the 
metric for a black hole that carries angular momentum. Although metric (\ref{BTZJ})
was obtained by a coordinate transformation from (\ref{BTZ}) and is therefore 
locally isometric to the latter (as all of our spaces are locally isometric to 
AdS space), it differs in its global structure: we have silently assumed that 
the new metric (\ref{BTZJ}) is periodic with period $2\pi$ in the {\it new} 
angular variable $\varphi$, rather than in the old variable $\phi$. This means,
in coordinate-independent language, that we have changed the identification
group that creates this new spacetime from AdS space. As for the non-rotating
BTZ black hole, the new group for this ``single" rotating black hole is still
generated by all the powers of a single isometry of AdS space, but this
isometry does not leave invariant a totally geodesic spacelike surface of time
symmetry. The surface $T =$ const that is obviously left invariant by a 
displacement of $\varphi$ is twisted, as measured by its extrinsic curvature,
and this is one indication of the global difference from the static metric.

When only the one new coordinate $\varphi$ changes by $2\pi$, the old 
coordinates of (\ref{BTZ}) change by
\begin{equation}
t\rightarrow t+{\pi J\over m} \qquad \phi \rightarrow \phi + 2\pi.
\label{twist}
\end{equation}
A change in either $t$ or $\phi$ is of course an isometry of the metric 
(\ref{BTZ}), and because $t$ and $\phi$ are coordinates, the two changes 
commute. The identification 
for a rotating black hole involves the two isometries applied simultaneously.
Either one is a ``boost" about an axis of fixed points;
the change in $\phi$ has fixed points in the future, at $r=0$, and the 
change in $t$ has fixed points at $r=\ell\sqrt{m}$, the horizon of 
(\ref{BTZ}).\footnote{Since these 
isometries are also isometries of the 
periodically identified embedding (\ref{ads}), each axis of fixed points is 
really repeated an infinite number of times in AdS space itself.}  The 
combination of the two does not have any fixed points at all (either one 
moves points on the fixed axis of the other in the direction of the axis): the 
length $R^2$ of the corresponding Killing vector $\partial/\partial\varphi$ 
vanishes where the vector is null but not zero,
since its scalar product with the finite $\partial/\partial T$ is the finite 
constant $J$. Earlier we argued 
that a spacelike set of fixed points of the identification isometry becomes a 
kind of singularity after the identification, and its removal from the 
spacetime gave us the end of \scri\ and associated horizon. What happens 
when we do not have this singularity?

\subsection{Is it a Black Hole?}

The geometry of metric (\ref{BTZJ}) --- more properly speaking, the 
geometry of its analytic extension, or of AdS space identified according 
to the $\varphi\rightarrow\varphi+2\pi$ isometry exhibited by this 
metric --- satisfies the definition of a black hole 
if we are somewhat creative about the definition of ``singularity." We 
expect the singularity to occur at $R=0$, but because there are no fixed 
points, the identified spacetime is regular there, and can be continued to 
negative $R^2$. But then the closed $\varphi$-direction becomes timelike, 
hence the spacetime has a region of closed timelike lines. We shall follow the 
usual practice to regard these as sufficiently unphysical that they should be 
eliminated from the spacetime, like a singularity. So we confine attention 
to $R>0$.

Our spacetime then ends at the singularity surfaces where the square of 
the Killing vector $\partial/\partial\varphi$ vanishes, $R^2=0$. The 
corresponding $r^2$ of Eq (\ref{coord}) is negative. We recall from 
Sect.~2.1 that this occurs on two timelike surfaces in a region where 
$\phi$ is timelike, unlike the non-rotating black hole whose singularity 
occurs on the spacelike line $r=0$. Since there is a singularity-free 
region between the two singularity surfaces, not all timelike lines that 
``fall into the black hole" (cross the horizon) end at the singularity; 
they can escape through the hole left open by the singularity surfaces, 
as is the case in a three-dimensional Kerr black hole.
However, at conformal infinity the difference between $R$ and $r$ 
disappears, the two singularity surfaces come together at the point where the 
spacelike line $r=0$ meets conformal infinity.\

Thus the covering space of \scri\ for the rotating black hole looks the 
same as that of the non-rotating one that corresponds to it via 
Eq (\ref{mass}), only the identification is different, as shown in Fig.~16b.
We see that \scri\ has an endpoint,
there is a horizon, so the identified spacetime is a black hole.

We can recognize a (rotating) black hole in a spacetime by the presence
of a closed, non-contractible spacelike geodesic $\gamma$. If we have 
such a $\gamma$ we consider all spacelike geodesics that start normal 
to $\gamma$. We assume that these can be divided into two types, which 
we might call right-starting and left-starting (with respect to an 
arbitrarily chosen direction of $\gamma$). If all geodesics of one type
reach infinity, then they cover the outside of a black hole. In this region
the totally geodesic timelike surfaces normal to $\gamma$ are surfaces
of constant $\phi$. Within these surfaces one can introduce
coordinates so that the metric takes the form (\ref{BTZJ}). (If the normals 
to those surfaces, not at $\gamma$, also integrate to closed curves 
after one circuit of $\gamma$, we have $J=0$.) 

\subsection{Does it rotate?}

The asymptotically measurable properties of (2+1)-dimensional black holes 
can be defined in various way, for example: from the ADM form of the Einstein 
action; as the conserved 
quantities that go with the Killing vectors $\partial/\partial t$ and 
$\partial/\partial\varphi$; as the Noether charges associated with $t$- 
and $\varphi$-displacements; and so on \cite{mass,BTZ}. All of these yield
$M$ as the mass and $J$ as the angular momentum. 

$J$ can also be measured ``quasi-locally" in the neighborhood of the 
horizon. We find an extremal closed spacelike geodesic (corresponding to 
$r=\ell$) and parallel transport an orthogonal vector around this 
geodesic. According to Eq (\ref{twist}) the hyperbolic ``holonomy" angle 
between the original and rotated vectors is $\pi J/m\ell$.
   
\subsection{Multiple Black Holes with Angular Momentum}

It is fairly straightforward to 
extend the methods of section 2.4 to obtain metrics with several asymptotic
regions, or with non-standard topologies, that have angular momentum as 
measured in these asymptotic regions; the main difference is that we will 
deal with spacetimes rather than initial values.
Our aim is only to show that rotating multi-black-holes are possible, and to 
indicate what the free parameters are.

We begin with a three-black-hole spacetime, whose time development can be
described by the geometry of Fig.~13. We suppose that the front 
left and right surfaces are identified, and similarly the back left and
right surfaces. The corresponding fixed points are the front and back
edges of the pyramid. As we have seen, there is then a third black hole
associated with the left and right edges (which are identified with each
other). We cut this figure into two halves by the plane $S$ (a totally geodesic
surface) spanned by these left and right edges. This surface cuts 
the third black hole
into two equal parts, which we can think of, for example, as $\phi = 0$ to
$\phi = \pi$ and $\phi = \pi$ to $\phi = 2\pi$, respectively. Now we 
re-identify the two halves with a ``boost" between them, 
that is an isometry with fixed points along the normal to the plane $S$
at the center of the initial surface, as illustrated in Fig.~17.

\begin{figure}
\unitlength 1.0mm
\linethickness{0.4pt}
\begin{picture}(86.00,68.00)(-7,10)
\bezier{30}(20.15,32.00)(20.15,38.59)(35.08,42.34)
\bezier{30}(64.92,42.34)(79.85,38.59)(79.85,32.00)
\bezier{30}(20.15,32.00)(20.15,25.41)(35.08,21.66)
\bezier{40}(35.08,21.66)(50.00,18.71)(64.92,21.66)
\bezier{30}(64.92,21.66)(79.85,25.41)(79.85,32.00)
\bezier{14}(34.17,60.00)(34.17,63.17)(42.09,64.97)
\bezier{20}(42.09,64.97)(50.00,66.39)(57.91,64.97)
\bezier{14}(57.91,64.97)(65.83,63.17)(65.83,60.00)
\bezier{14}(34.17,60.00)(34.17,56.83)(42.09,55.03)
\bezier{20}(42.09,55.03)(50.00,53.61)(57.91,55.03)
\bezier{14}(57.91,55.03)(65.83,56.83)(65.83,60.00)
\bezier{264}(24.00,35.00)(50.00,55.00)(76.00,35.00)
\put(54.00,68.00){\vector(-1,-2){3.50}}
\put(55.00,68.00){\makebox(0,0)[lc]{C (Center of projection)}}
\put(29.00,65.00){\vector(1,-1){4.50}}
\put(28.50,65.00){\makebox(0,0)[rc]{Conformal infinity}}
\bezier{36}(23.86,35.09)(21.75,31.75)(24.74,28.25)
\bezier{92}(38.25,22.63)(49.82,20.88)(61.40,22.63)
\bezier{60}(24.91,28.07)(28.77,24.39)(38.42,22.63)
\bezier{36}(76.14,35.09)(78.25,31.75)(75.26,28.25)
\bezier{60}(75.09,28.07)(71.23,24.39)(61.58,22.63)
\bezier{26}(34.04,60.00)(34.04,50.35)(25.09,39.30)
\put(71.93,47.37){\vector(-3,-2){9.65}}
\put(72.28,47.37){\makebox(0,0)[lc]{Initial surface}}
\bezier{26}(65.96,60.00)(65.96,50.35)(74.91,39.30)
\thicklines
\put(44.00,56.00){\line(6,1){12.00}}
\put(44.00,56.00){\line(1,-6){6.33}}
\bezier{20}(43.00,62.00)(42.30,61.20)(36.00,54.00)
\bezier{14}(42.90,62.00)(43.05,61.20)(43.85,56.00)
\bezier{18}(43.00,62.00)(43.70,61.50)(50.00,57.00)
\bezier{18}(57.00,62.00)(56.30,61.50)(50.00,57.00)
\bezier{20}(57.00,62.00)(57.70,61.30)(64.00,55.00)
\put(50.00,60.00){\circle*{0.00}}
\put(56.00,58.00){\line(6,-5){30.00}}
\put(85.9,32.58){\line(-5,-2){36.00}}
\put(44.00,56.00){\line(-6,-5){28.50}}
\put(15.25,32.1){\line(5,-2){34.50}}
\thinlines
\bezier{10}(57.00,62.00)(56.90,61.60)(56.00,58.00)
\put(56.00,58.00){\line(-1,-4){3.25}}
\put(86.00,33.00){\line(-5,2){16.00}}
\put(15.50,32.20){\line(5,2){10.00}}
\end{picture}

\begin{caption}
{A three-black-hole geometry obtained by cutting Fig.~13 into two
tetrahedra by the plane $S$ of the paper (passing through C), 
and re-gluing after an isometry with axis normal to that plane. 
The isometry moves the top of the front tetrahedron from C
to the left (and up), and the top of the back tetrahedron 
from C to the right (and up). The dotted outlines show these two tops. 
The solid figure approximates the convex region bounded by the four planes 
that are 
identified pairwise (but it is {\em not} the fundamental domain). The
edges where the planes intersect are drawn only to identify these planes;
they are simplified as straight lines (but ought to be hyperbolic arcs, 
representing geodesics). Unlike in Fig.~13 the edges are not to be
considered as singularities, except for the front and back edges, which are
fixed points of the two basic identifications that generate this 
spacetime. 
The other ``singularities" are the boundaries of the regions of closed
timelike lines, not drawn (and not easily identified) in this figure.}
\end{caption}
\end{figure}

The four planes stick out of the conformal infinity surface at the four
bottom corners, uncovering four parts of conformal infinity. As in Fig.~13,
the left and right infinity parts combine into one continuous region due to the
identifications. So this spacetime has three conformal infinities with ends, 
and therefore represents three black holes.

The two black holes associated with the front and back edges, as seen from
their respective asymptotic regions, are unchanged by this re-identification: 
by a ``boost" isometry either of these edges and
associated planes (but not both together) can be moved back to their old position.
Since the two planes alone determine the asymptotic behavior of the black hole,
either of these holes has the same mass, and vanishing angular momentum, as
before. But the third black hole changes, because the left and
right edges no longer lie in the same plane. As we go once around this third black
hole, we cross the surface $S$ twice, and its effects add (as a right-handed screw 
is right-handed from either end). The black hole therefore acquires angular momentum.
Unfortunately this is not directly described by Eqs (19-22), because the ``boost" in
Eq (\ref{twist}) has fixed points at the horizon of the non-rotating black hole,
whereas the fixed points of the boost of Fig.~17 lie along a geodesic connecting
the asymptotic regions of the two other holes. However, for the third black hole
this difference is asymptotically negligible: as seen from its own infinity it
does have angular momentum. (Its
standard form (\ref{BTZJ}) would correspond to identification surfaces different
from any of those drawn in Fig.~17.)

By a similar re-identification any one of a $k$-black-hole time-symmetric
spacetime can be given angular momentum; further momentum parameters will be
needed to describe how the asymptotic regions fit to an interior.
Generally we expect one momentum parameter for each configuration
parameter of the corresponding time-symmetric spacetime. For example,
the toroidal black hole constructed as in Fig.~12 should allow three
independent momenta. Of these the state in which there is angular momentum
of the black hole as seen from infinity has been constructed \cite{ABH}. 
(Another state with momentum can be obtained from Fig.~17 by identifying
opposite rather than adjacent planes.)

\section{Conclusions}
We have seen that a considerable variety of black hole and 
multiply-connected spacetimes can be constructed by cutting a region out 
of anti-de Sitter space and identifying the cuts in various ways. Many of 
the properties, such as horizon structure and topological 
features of the time-symmetric spacetimes, have been investigated in 
detail. Comparatively little beyond existence is known about the 
spacetimes with angular momentum (but see \cite{ABH}).

\section*{Acknowledgment}
This research was supported in part by the National Science Foundation under Grant
No.~PHY94-07194.

\end{document}